\documentclass{SciPost}
\hypersetup{
  colorlinks,
  linkcolor={red!50!black},
  citecolor={blue!50!black},
  urlcolor={blue!80!black}
}
\usepackage[bitstream-charter]{mathdesign}
\DeclareSymbolFont{usualmathcal}{OMS}{cmsy}{m}{n}
\DeclareSymbolFontAlphabet{\mathcal}{usualmathcal}
\usepackage{comment}
\usepackage{amsmath}
\usepackage{amsthm}
\usepackage{bm}
\usepackage{braket}
\numberwithin{equation}{section}

\DeclareMathOperator{\ch}{ch}
\DeclareMathOperator{\Tr}{Tr}

\usepackage{graphicx}
\usepackage{subcaption}
\usepackage{xcolor}
\usepackage{enumitem}

\usepackage{tikz}
\usetikzlibrary{
  graphs,
  patterns,
  decorations.markings,
  decorations.pathreplacing,
  arrows.meta,
  matrix,
  shapes.misc,
  shapes.geometric
}

\usepackage{cite}

\theoremstyle{plain}

\theoremstyle{remark}

\newcommand{\order}[1]{\mathcal{O}\!\left(#1\right)}

\newcommand{\eq}[1]{\begin{equation}\begin{split}#1 \end{split}\end{equation}}
\newcommand{\eqnn}[1]{\begin{equation*}\begin{split}#1 \end{split}\end{equation*}}
\newcommand{\eqs}[1]{\begin{align} #1 \end{align}}
\newcommand{\eqsnn}[1]{\begin{align*} #1 \end{align*}}

\newcommand{\dd}{\mathrm d}
\newcommand{\xx}{\mathrm{XX}}
\newcommand{\ind}[1]{\chi\!\left[#1\right]}

\fancypagestyle{SPstyle}{
\fancyhf{}
\lhead{\colorbox{scipostblue}{\bf \color{white} ~SciPost Physics }}
\rhead{{\bf \color{scipostdeepblue} ~Submission }}

\fancyfoot[C]{\textbf{\thepage}}
}
\begin{document}
\pagestyle{SPstyle}

\begin{center}{\Large \textbf{\color{scipostdeepblue}{
Exact anomalous current fluctuations in the strong-anisotropy limit of the XXZ chain
}}}\end{center}
\begin{center}\textbf{
Taiki Ishiyama\textsuperscript{1$\star$},
Taiga Kurose\textsuperscript{1},
Kazuya Fujimoto\textsuperscript{1,2}
and Tomohiro Sasamoto\textsuperscript{1}
}\end{center}
\begin{center}
{\bf 1} Department of Physics, Institute of Science Tokyo,
2-12-1 Ookayama, Meguro-ku, Tokyo 152-8551, Japan\\
{\bf 2} Institute for Liberal Arts, Institute of Science Tokyo,
Ichikawa, 272-0827, Japan
\\[\baselineskip]
$\star$ \href{mailto:ishiyama.t.d861@m.isct.ac.jp}{\small ishiyama.t.d861@m.isct.ac.jp}
\end{center}

\section*{\color{scipostdeepblue}{Abstract}}
\textbf{\boldmath{%
We study spin-current fluctuations of the folded XXZ chain, which describes the strong-anisotropy limit of the XXZ chain. 
We obtain an exact expression for the moment-generating function of the time-integrated spin current at infinite temperature. 
A long-time asymptotic analysis establishes that the rescaled time-integrated current
converges to a non-Gaussian distribution described by the M-Wright function.
The derivation relies on a nonlocal map to the one-dimensional $t$--$0$ model and its exact spin--charge separation.
By extending this approach, we also show that the dynamical correlation function of the spins
summed over two adjacent sites exhibits a diffusive Gaussian profile at long times.
}}
\vspace{\baselineskip}

\noindent\textcolor{white!90!black}{%
\fbox{\parbox{0.975\linewidth}{%
\textcolor{white!40!black}{\begin{tabular}{lr}%
  \begin{minipage}{0.6\textwidth}%
    {\small Copyright attribution to authors. \newline
    This work is a submission to SciPost Physics. \newline
    License information to appear upon publication. \newline
    Publication information to appear upon publication.}
  \end{minipage} & \begin{minipage}{0.4\textwidth}
    {\small Received Date \newline Accepted Date \newline Published Date}%
  \end{minipage}
\end{tabular}}
}}
}

\vspace{10pt}
\noindent\rule{\textwidth}{1pt}
\tableofcontents
\noindent\rule{\textwidth}{1pt}
\vspace{10pt}

\section{Introduction}
\label{sec:introduction}
Understanding universal aspects of transport in interacting many-body systems is a central problem in statistical mechanics.
In classical systems, the time-integrated current, namely the net amount of a conserved quantity
transported across a given point up to time $t$, has served as a fundamental probe of transport dynamics.
Its statistics characterize not only transport behavior, such as ballistic or diffusive transport,
but also universal properties of dynamical fluctuations.
A prominent example is the Kardar--Parisi--Zhang universality class~\cite{Corwin_2012, Takeuchi_2018},
in which systems exhibit universal non-Gaussian current fluctuations.

Recently, there has been growing theoretical interest in current fluctuations
in quantum many-body systems as a means of exploring universal dynamics.
This interest has been particularly driven by findings that, in thermal equilibrium,
isotropic integrable systems such as the XXX spin chain
exhibit behavior consistent with the KPZ universality class
in the variance of the time-integrated current
as well as in two-point dynamical correlation functions~\cite{Ljubotina_2019,Das_2019,Krajnik_2020,Fava_2020,Ye_2022,Roy_2023,Moca_2023,Takeuchi_2025,Moca_2026_2}.
Remarkably, however, higher-order current fluctuations in these systems deviate from the KPZ predictions,
motivating the study of the probability distribution of the time-integrated current itself.
Furthermore, rapid experimental progress in cold-atom systems and superconducting quantum circuits has stimulated research in this direction,
as the statistics of time-integrated currents have become experimentally accessible~\cite{Wei_2022,Wienand_2024,Google, Kwon_2026}.

Against this background of growing interest in the full current distribution, 
one of the most intriguing findings is the emergence
of anomalous current fluctuations in the easy-axis XXZ spin chain~\cite{Krajnik_2022_2,Gopalakrishnan_2024,Yoshimura_2026}.
In particular, the time-integrated spin current at thermal equilibrium exhibits a non-Gaussian distribution described by the M-Wright function.
This distribution was first obtained through an exact analysis of a classical automaton~\cite{Krajnik_2022}, 
and was subsequently found in several classical and quantum systems~\cite{Kormos_2022,Gopalakrishnan_2024_2,Krajnik_2024,Krajnik_2024_2,Krajnik_2025,
Pozsgay_2026, Moca_2026,Patu_2026,Moca_2026_3},
suggesting that such current fluctuations are universal.
The origin of these anomalous fluctuations has been clarified in several tractable settings.
In simple classical automata and quantum models,
exact microscopic calculations as well as hydrodynamic approaches have revealed
that the non-Gaussian distribution can be understood as arising from
the interplay of two underlying Gaussian fluctuations~\cite{Krajnik_2022,Yoshimura_2025,Fujimoto_2026}.
For the XXZ spin chain, this mechanism was first studied through
a phenomenological argument in the strong-anisotropy limit~\cite{Gopalakrishnan_2024}, 
and was subsequently established at the hydrodynamic level~\cite{Yoshimura_2026}.
However, an exact microscopic derivation of the anomalous current fluctuations
has yet to be achieved, even in the strong-anisotropy limit.

In this work, we consider the folded XXZ chain~\cite{Yang_2020, Pozsgay_2021_1, Zadnik_2021_1, Zadnik_2021_2}, 
which describes the strong-anisotropy limit of the XXZ spin chain.
We derive an exact expression for the moment-generating function (MGF)
of the time-integrated spin current at infinite temperature.
By performing a long-time asymptotic analysis of this exact expression,
we show that the time-integrated spin current follows the non-Gaussian distribution described by the M-Wright function.
The exact solution is obtained through a mapping to the one-dimensional $t$--$0$ model~\cite{Dias_2000,Pozsgay_2021_1,Feldmeier_2022},
which arises as an effective model of the Hubbard model in the limit of infinite repulsion~\cite{Essler_hubbard}.
The $t$--$0$ model exhibits exact spin--charge separation~\cite{Ogata_1990, Izergin_1998, Gamayun_2023, Gamayun_2024}:
the spin configuration remains static, while the charge dynamics is equivalent to that of noninteracting fermions.
Exploiting this structure, we identify microscopically the two Gaussian fluctuations underlying the non-Gaussian current distribution.
Furthermore, we apply our method to the dynamical correlation function of spins summed over two adjacent sites, 
deriving a diffusive Gaussian profile in the long-time limit.

We remark that, in the $t$--$0$ model, an exact microscopic derivation
of anomalous spin-current fluctuations has already been obtained~\cite{Fujimoto_2026}.
However, the mapping between the folded XXZ and $t$--$0$ models is nonlocal,
and the physical spin current in the folded XXZ model is not mapped directly onto the spin current of the $t$--$0$ model.
Consequently, the existing solution cannot be applied directly to the present problem,
and an additional treatment of the nonlocal relation between the two models is required.
We also note that Ref.~\cite{Pozsgay_2026} has recently studied related current fluctuations in the stochastic XNOR hopping model, 
a stochastic analogue of the folded XXZ chain~\cite{Singh_2021,Feldmeier_2022,Gopalakrishnan_2024}.
In that setting, however, no exact finite-time expression for the current statistics was obtained,
and the long-time asymptotic form was proposed as a conjecture.
Our result instead provides an exact finite-time MGF for the quantum folded XXZ model
and an analytical derivation of its long-time asymptotics.

The paper is organized as follows.
Section~\ref{sec:setup} defines the model and the time-integrated spin current,
and Sec.~\ref{sec:main-results} presents our main results.
Section~\ref{sec:mapping} introduces the mapping to the $t$--$0$ model,
which is used in Sec.~\ref{sec:exact-mgf} to derive the exact MGF.
Section~\ref{sec:long-time-asymptotics} gives its long-time asymptotic analysis.
Section~\ref{sec:spin-correlation} extends our approach to the dynamical correlation function of spins.
Finally, Sec.~\ref{sec:conclusion} concludes the paper.

\section{Setup}
\label{sec:setup}
We consider a one-dimensional spin-$1/2$ system on the infinite lattice $\mathbb{Z}$. 
The local Hilbert space at each site $j\in\mathbb{Z}$ is $\mathcal H_j\simeq\mathbb C^2$. 
The time evolution is generated by the folded XXZ Hamiltonian,
\eq{
H=-\sum_{j\in \mathbb{Z}}(s_j^+s_{j+1}^-+s_{j+1}^+s_j^-)\left(s_{j-1}^zs_{j+2}^z+\frac14\right),
}
where $s_j^+$, $s_j^-$, and $s_j^z$ denote the spin-$1/2$ raising, lowering, 
and $z$-component operators at site $j$, respectively.
Each local term in the Hamiltonian acts on four consecutive sites and exchanges
the two inner spins only when the two outer spins are parallel.
The resulting constrained dynamics conserves the total number of domain walls,
$
\sum_{j\in \mathbb{Z}}(1/2 -2s_j^zs_{j+1}^z ).
$
Since only spin-exchange processes that preserve the number of domain walls are allowed in the infinite-anisotropy 
limit of the XXZ spin chain, 
the folded XXZ model describes the effective dynamics in this limit~\cite{Yang_2020, Pozsgay_2021_1, Zadnik_2021_1, Zadnik_2021_2}.

In this work, we consider the infinite-temperature initial state,
\eq{
\rho=\bigotimes_{j\in \mathbb{Z}} \left[\frac12\ket{\uparrow}\bra{\uparrow}+\frac12\ket{\downarrow}\bra{\downarrow}\right]_j,
\label{eq:initial}
}
where $\ket{\uparrow}$ and $\ket{\downarrow}$ are the eigenstates of $s^z$ with eigenvalues $1/2$ and $-1/2$, respectively.

The quantity of interest in this work is the probability distribution of the time-integrated spin current $Q_t$.
Here, $Q_t$ represents the total magnetization transferred from right to left across 
the bond between sites $0$ and $1$ during the time interval $[0,t]$. 
We define $Q_t$ through two projective measurements of the magnetization in the left subsystem~\cite{Esposito_2009},
\eq{
S_{\leq 0}:=\sum_{j \leq 0} s_j^z,
}
performed at times $0$ and $t$. 
If the corresponding measurement outcomes are $m$ and $m+n$, respectively, the time-integrated current is $Q_t=n$.
Let $P_m$ denote the projection operator onto the eigenspace of $S_{\leq 0}$ with eigenvalue $m$. 
The probability distribution of $Q_t$ is then given by~\cite{Esposito_2009}
\eq{
\Pr[Q_t=n]=\sum_m\Tr\left[P_{m+n}e^{-iHt}P_m\rho P_m e^{iHt}\right].
}
Rather than directly analyzing this probability distribution, it is convenient to consider the MGF,
\eq{
\braket{e^{\lambda Q_t}}:=\sum_{n\in\mathbb Z}e^{\lambda n}\Pr[Q_t=n].
}
Since the infinite-temperature initial state satisfies $[\rho,S_{\leq 0}]=0$, the MGF can be written in the compact form
\eq{
\braket{e^{\lambda Q_t}} = \Tr\left[e^{\lambda S_{\leq 0}}e^{-iHt}e^{-\lambda S_{\leq 0}}\rho e^{iHt}\right].
\label{eq:def-mgf}
}
In what follows, we derive the exact expression for the MGF and analyze its long-time asymptotic behavior.

\section{Main results}
\label{sec:main-results}

We summarize here the two main results of this work.
The first is an exact finite-time expression for the MGF. The second is its long-time scaling limit.

\subsection{Exact expression for the MGF}

The exact MGF is
\eq{
\braket{e^{\lambda Q_t}}
&=1+\frac43\sum_{\epsilon=0}^{1}\sum_{n=1}^{\infty}
\sum_{k=\epsilon}^{n}B_{n,k}\,
\mathcal P_t^{(\epsilon+\epsilon'_\epsilon)}
\left(x_\epsilon(k)-\epsilon,n\right)
\left\{\mathcal G_\lambda^{(\epsilon)}(n,k)-1\right\}.
\label{eq:main-exact-mgf}
}
Here,
\eq{
\epsilon'_\epsilon:=(\epsilon+k)\bmod2,\qquad
x_\epsilon(k):=\frac{k}{2}+\frac{(-1)^\epsilon}{4}\left[(-1)^k-1\right],
\qquad \epsilon=0,1,
\label{eq:def-x}
}
and
\eq{
B_{n,k}:=\binom nk\left(\frac23\right)^k\left(\frac13\right)^{n-k}.
\label{eq:binom-B}
}
The quantity $\mathcal G_\lambda^{(\epsilon)}(n,k)$ appearing in Eq.~\eqref{eq:main-exact-mgf} is
\eq{
\mathcal G_{\lambda}^{(\epsilon)}(n,k)
:=\sum_{A=0}^{k}W_{n,k}(A)
\ch\left[\lambda\left\{A-\frac{k}{2}+d_\epsilon(n,k)\right\}\right],
\label{eq:main-Gcal}
}
where
\eq{
d_\epsilon(n,k)
:=\frac14(-1)^{n-k}\left[1+(-1)^{\epsilon+k}\right]
-\frac14\left[1+(-1)^\epsilon\right],
}
and
\eq{
W_{n,k}(A) :=
\frac{1}{\binom nk}
\binom{A+\lfloor(n-k)/2\rfloor}{\lfloor(n-k)/2\rfloor}
\binom{k-A+\lceil(n-k)/2\rceil-1}{\lceil(n-k)/2\rceil-1}.
\label{eq:main-W}
}
For $n=k$, we use the convention
$\binom{-1}{-1}=1$ and $\binom{r}{-1}=0$ for $r\neq-1$.
The function $\mathcal P_t^{(a)}(x,n)$ in Eq.~\eqref{eq:main-exact-mgf}
is defined for $a=0,1,2$, $x\in\mathbb Z_{\geq0}$, and $n\in\mathbb Z$ by
\eq{
\mathcal P_t^{(a)}(x,n)
:=
\int_{-\pi}^{\pi}\frac{d\theta}{2\pi}e^{-in\theta}
\bigl(\mathsf T_\theta^a F_t(\theta;\cdot)\bigr)(x),
\label{eq:main-pt}
}
where $\mathsf T_\theta$ is the difference operator acting on the spatial argument,
\eq{
(\mathsf T_\theta f)(x)
:=\frac{(3e^{i\theta}-1)f(x)-2f(x+1)}{3(e^{i\theta}-1)}.
\label{eq:main-difference-operator}
}
The function $F_t$ is defined by
\eq{
F_t(\theta;x)
:=g(\theta)^x
\det\left[1+\omega(\theta)K_{\mathrm{Bes}}^{(x)}\right]_{L^2([0,t^2])},
\label{eq:main-charge-determinant}
}
with
\eq{
g(\theta):=\frac{1+3e^{i\theta}}{4},
\qquad
\omega(\theta):=\frac{3}{16}\left(e^{i\theta}+e^{-i\theta}-2\right).
\label{eq:main-g-omega}
}
The determinant in Eq.~\eqref{eq:main-charge-determinant}
is a Fredholm determinant on $L^2([0,t^2])$
with the continuous Bessel kernel
\eq{
K_{\mathrm{Bes}}^{(x)}(u,v) :=
\frac{
J_x(\sqrt u)\sqrt v\,J_x'(\sqrt v)
-\sqrt u\,J_x'(\sqrt u)J_x(\sqrt v)
}{2(u-v)}.
\label{eq:main-continuous-bessel}
}
Here, $J_x$ is the Bessel function of the first kind of order $x$,
and the prime denotes differentiation with respect to its argument.

Equations~\eqref{eq:main-exact-mgf}--\eqref{eq:main-continuous-bessel}
constitute the exact finite-time result for the infinite system.
Their derivation is given in Sec.~\ref{sec:exact-mgf} and the appendices.

The advantage of Eq.~\eqref{eq:main-exact-mgf} is that the current MGF is expressed in terms of two tractable ingredients: 
a combinatorial factor and a determinant associated with free fermions.
This representation makes the long-time asymptotic analysis particularly accessible.

The origin of this structure is the exact mapping of the folded XXZ spin chain to the one-dimensional $t$--$0$ model~\cite{Pozsgay_2021_1}, 
which is the effective model of the Hubbard model in the limit of infinitely strong repulsive interaction~\cite{Essler_hubbard}.
The $t$--$0$ model exhibits an exact spin--charge separation~\cite{Ogata_1990, Izergin_1998, Gamayun_2023, Gamayun_2024}: 
the spin degrees of freedom are static, whereas the charge degrees of freedom evolve according to the XX spin chain.
Accordingly, $\mathcal G_\lambda^{(\epsilon)}(n,k)$ is obtained by averaging
over the static spin labels, while $\mathcal P_t^{(a)}(x,n)$ is determined by the XX-chain dynamics.
Thus, the original interacting folded XXZ problem is reduced to a combination of a static combinatorial problem and a free-fermion dynamical problem.

We remark that a related decomposition has been obtained
for the spin-current fluctuations of the $t$--$0$ model itself in Ref.~\cite{Fujimoto_2026}.
However, the mapping between the folded XXZ and $t$--$0$ models
is nonlocal, and the physical spins of the folded XXZ model
do not correspond directly to the $t$--$0$ spin labels.
Consequently, the folded XXZ spin current is a different observable
from the $t$--$0$ spin current, and deriving its MGF requires
an explicit treatment of this nonlocal mapping.

\subsection{Long-time behavior of current fluctuations}

In the long-time limit, the time-integrated current fluctuates on the scale $Q_t\sim t^{1/4}$.
The limiting MGF is
\eq{
\lim_{t\to\infty}\left\langle e^{\lambda Q_t/t^{1/4}}\right\rangle =
\sqrt3\int_0^\infty\dd y\,
\exp\left(-\frac{3\pi}{4}y^2+\frac{\lambda^2}{4}y\right).
\label{eq:main-asymptotic-mgf}
}
As derived in Sec.~\ref{sec:long-time-asymptotics},
this limiting MGF arises from the interplay of two Gaussian fluctuations in the $t$--$0$ spin and charge contributions,
$\mathcal G_\lambda^{(\epsilon)}$ and $\mathcal P_t^{(a)}$, respectively.
This provides a microscopic derivation of the anomalous spin-current fluctuations in the folded XXZ model.

The limiting probability density of the rescaled current $q:=Q_t/t^{1/4}$ is obtained
by inverting Eq.~\eqref{eq:main-asymptotic-mgf}.
To express the result, we define the M-Wright function
\eq{
p_{\mathrm{MW}}(q;\sigma) := \frac{1}{\pi\sigma} \int_0^\infty\frac{\dd y}{\sqrt{y}}\,
\exp\left(-\frac{y^2}{2\sigma^2}-\frac{q^2}{2y}\right).
\label{eq:def-MW}
}
We then obtain
\eq{
\lim_{t\to\infty} \Pr\left[\frac{Q_t}{t^{1/4}}\leq q\right] =
\int_{-\infty}^{q}\dd q'\, p_{\mathrm{MW}}\left(q';\frac{1}{\sqrt{6\pi}}\right).
\label{eq:main-limiting-density}
}
Thus, the limiting current distribution is non-Gaussian and is described by the M-Wright function.
We remark that this distribution is also referred to as the nested Gaussian distribution in the literature~\cite{Yoshimura_2026}.

\begin{figure}[t]
\centering
\includegraphics[width=0.9\linewidth]{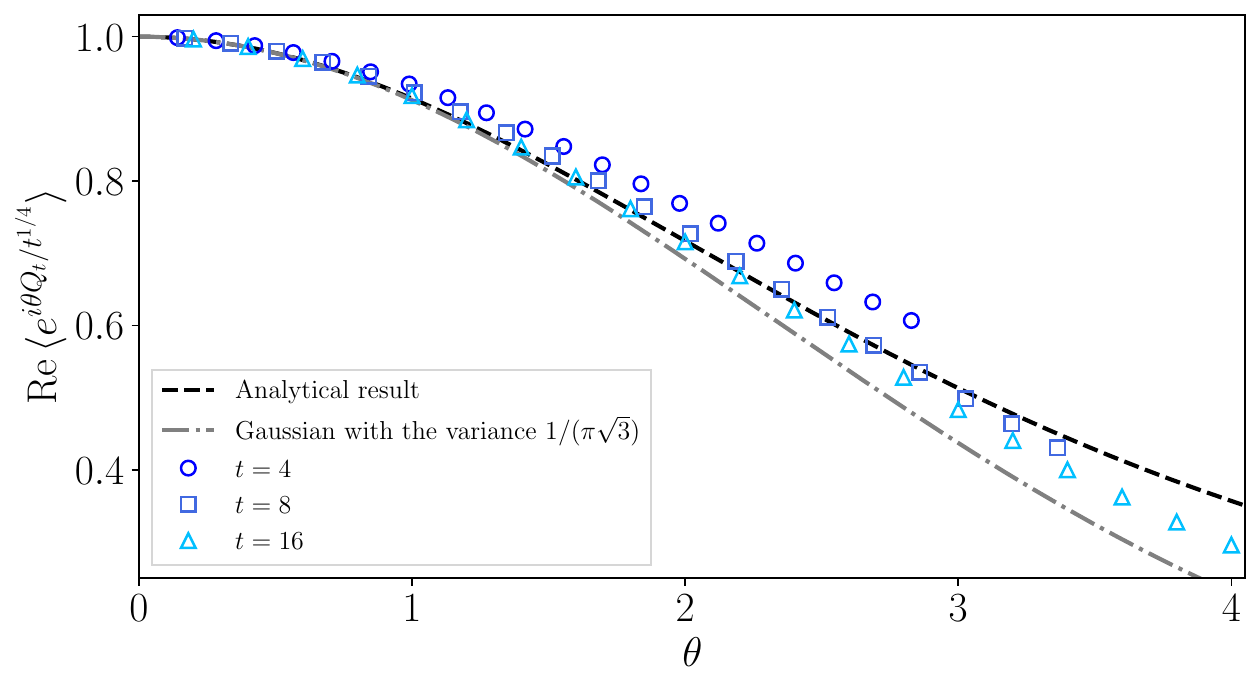}
\caption{
Real part of the characteristic function of the rescaled current, $\braket{e^{i\theta Q_t/t^{1/4}}}$.
Numerical results at $t=4$, $8$, and $16$ are compared with the long-time
prediction in Eq.~\eqref{eq:main-asymptotic-mgf}
and the characteristic function of a zero-mean Gaussian
with the same variance $\lim_{t \to \infty}\braket{Q^2_t}_c /t^{1/2}= 1/(\pi\sqrt3)$.
The numerical results are obtained from matrix product operator (MPO)
simulations~\cite{Schollwock_2011, Paeckel_2019,Valli_2025} of an integrable Trotterized circuit of the folded XXZ model~\cite{Pozsgay_2021_2,Gombor_2021},
with a system size $L=128$, a time-step parameter $u=0.1$, and a maximum MPO bond dimension $\chi=128$.
}
\label{fig:current-scaling}
\end{figure}

We compare the limiting MGF in Eq.~\eqref{eq:main-asymptotic-mgf}, evaluated at $\lambda=i\theta$,
with a numerical calculation of the rescaled characteristic function
$\braket{e^{i\theta Q_t/t^{1/4}}}$ under the folded XXZ dynamics.
See Appendix~\ref{app:numerics} for the details of the numerical method and convergence checks.
Figure~\ref{fig:current-scaling} shows its real part as a function of $\theta$ at different times,
together with the long-time prediction obtained by setting $\lambda=i\theta$ in Eq.~\eqref{eq:main-asymptotic-mgf}.
For comparison, we also plot the characteristic function of a zero-mean Gaussian with the same asymptotic variance,
$\lim_{t\to\infty}\braket{Q_t^2}_c/t^{1/2}=1/(\pi\sqrt{3})$.
As shown in Fig.~\ref{fig:current-scaling}, for $\theta\lesssim2$,
the $t=16$ data are in excellent agreement with the asymptotic result and are distinguishable from the Gaussian prediction.
For $\theta>2$, the numerical data at $t=16$ show the visible deviations from the asymptotic result.
A related issue of slow convergence near the central cusp of the current probability distribution has been observed
in the $t$--$0$ model~\cite{Fujimoto_2026}.
Under the Fourier transform, this cusp corresponds to the algebraic decay of the limiting characteristic function,
$\lim_{t\to\infty}\braket{e^{i\theta Q_t/t^{1/4}}} \simeq 4\sqrt{3}/\theta^2$ as $|\theta|\to\infty$.
The finite-time deviations observed here at larger $\theta$
are therefore consistent with the slow convergence reported in that study.
We expect these deviations to diminish at longer times.

A hydrodynamic analysis in Ref.~\cite{Yoshimura_2026} showed that,
at zero magnetization and finite temperature, the time-integrated spin current
of the easy-axis XXZ chain follows the same distribution described by the M-Wright function.
In the parametrization of Ref.~\cite{Yoshimura_2026}, $\sigma$ in Eq.~\eqref{eq:def-MW} satisfies 
$\sigma^2=C_s^2D_s$, where $C_s$ and $D_s$ denote the static
spin susceptibility and the spin diffusion constant, respectively.
At infinite temperature, $C_s=1/4$.
The convention for the diffusion constant in Ref.~\cite{Yoshimura_2026}
differs by a factor of two from that used in
Refs.~\cite{Gopalakrishnan_2019,Feldmeier_2022}, which give
$D=4/(3\pi)$ for the folded XXZ spin chain at infinite temperature.
Thus, in the convention of Ref.~\cite{Yoshimura_2026}, the corresponding
diffusion constant is $D_s=2D=8/(3\pi)$.
Substituting these values gives
\eq{
\sigma^2=C_s^2D_s
=
\left(\frac14\right)^2\frac{8}{3\pi}
=
\frac{1}{6\pi},
}
which agrees exactly with Eq.~\eqref{eq:main-limiting-density}.

\section[Mapping to the t--0 model]{Mapping to the $t$--$0$ model}
\label{sec:mapping}

In this section, we map the folded XXZ model to the $t$--$0$ model~\cite{Dias_2000, Pozsgay_2021_1, Feldmeier_2022}.
For this purpose, we first consider the model on the half-infinite
lattice $\mathbb Z_{\geq0}$, so that the mapping can be defined
by reading the configuration from the left boundary.
We use $H$ and $\rho$ to denote the Hamiltonian and the initial density matrix restricted to this lattice.
In particular, the Hamiltonian on $\mathbb{Z}_{\geq0}$ is given by
\eq{
H = - \sum_{j \geq 1}(s_j^+s_{j+1}^-+s_{j+1}^+s_j^-)\left(s_{j-1}^zs_{j+2}^z+\frac14\right).
\label{eq:H-half}
}

We place the measurement cut between sites $R$ and $R+1$
and define the magnetization
\eq{
S_R:=\sum_{j=0}^{R}s_j^z.
\label{eq:def-SR}
}
Throughout the following calculation, we take $R\in2\mathbb N$.
The current MGF in the half-infinite system is
\eq{
\braket{e^{\lambda Q_t}}_R := \Tr\left[
e^{\lambda S_R}e^{-iHt}e^{-\lambda S_R}\rho e^{iHt}
\right].
\label{eq:half-infinite-mgf}
}
The infinite-system MGF in Eq.~\eqref{eq:def-mgf} is recovered
by taking $R\to\infty$ at fixed $t$,
\eq{
\braket{e^{\lambda Q_t}} = \lim_{R\to\infty}\braket{e^{\lambda Q_t}}_R.
\label{eq:mgf-bulk-limit}
}

We first transform the folded XXZ model into a bond model and then map
the resulting bond dynamics to the $t$--$0$ model, following Ref.~\cite{Pozsgay_2021_1}.
After that, we derive an explicit representation of $S_R$ in the $t$--$0$ model,
which will be used in the calculation of the MGF.

\subsection{Bond model}

We first rewrite the folded XXZ model in terms of bond variables. 
For a spin configuration
\eq{
\bm{\sigma} :=(\sigma_j)_{j\geq0}\in\{\pm1\}^{\mathbb Z_{\geq0}},
}
where $+1$ and $-1$ correspond to $\uparrow$ and $\downarrow$, we define~\cite{Pozsgay_2021_1}
\eq{
b_j:=\frac12\left(1-\sigma_{j-1}\sigma_j\right),\qquad j\geq1,
\label{eq:bond-map}
}
and retain the boundary spin $\sigma:=\sigma_0$. 
The bond variable $b_j$ records whether the neighboring spins at sites $j-1$ and $j$ are parallel or antiparallel. 
The transformation
\eq{
\bm{\sigma}\longleftrightarrow(\bm b;\sigma)
}
is bijective, with inverse
\eq{
\sigma_j=\sigma(-1)^{\sum_{k=1}^{j}b_k}.
\label{eq:inverse-bond-transformation}
}

In the bond representation, the dynamics takes a particularly simple local form~\cite{Zadnik_2021_1,Pozsgay_2021_1}. 
Under the evolution of the folded Hamiltonian in Eq.~\eqref{eq:H-half}, 
the boundary spin $\sigma$ is conserved, and the only nontrivial transitions are 
\eq{
\uparrow \downarrow \uparrow \uparrow \; \longleftrightarrow \; \uparrow \uparrow \downarrow  \uparrow
} 
and its spin-reversed counterpart. Hence, the only allowed process in the bond variables is
\eq{
110\longleftrightarrow011.
}
Accordingly, the Hamiltonian \eqref{eq:H-half} in the bond representation is
\eq{
H_b=-\frac12\sum_{j=1}^{\infty} \left[ \ket{110}\bra{011}+\ket{011}\bra{110} \right]_{j,j+1,j+2}.
\label{eq:bond-Hamiltonian}
}
In addition, the infinite-temperature state \eqref{eq:initial}
remains a product state in the bond representation,
\eq{
\rho_b= \bigotimes_{j\geq1} \left[ \frac12\ket{0}\bra{0}+\frac12\ket{1}\bra{1} \right]_j.
\label{eq:bond-infinite-temperature-state}
}
For a fixed value of the conserved boundary spin $\sigma_0=\sigma$, 
the magnetization in the interval $[0,R]$ in Eq.~\eqref{eq:def-SR} is expressed as
\eq{
S_R^{(\sigma)} = \frac{\sigma}{2} + \frac{\sigma}{2}\sum_{j=1}^{R}(-1)^{\sum_{k=1}^{j}b_k}.
\label{eq:SR-bond}
}


Thus, the folded XXZ dynamics reduces to a constrained hopping dynamics in the bond variables, 
while the magnetization $S_R$ becomes nonlocal in this representation.
This bond description provides the starting point for the mapping to the $t$--$0$ model in the next subsection.

\subsection[t--0 model]{$t$--$0$ model}

The only nontrivial local process of the bond dynamics is $110\longleftrightarrow011$.
Moreover, the parity of the number of $1$'s between neighboring zeros is conserved under the dynamics.
These properties motivate the following mapping.
Reading the bond configuration from the left, we identify~\cite{Pozsgay_2021_1}
\eq{
0\longrightarrow\bullet_1,\qquad 10\longrightarrow\bullet_2,\qquad 11\longrightarrow\circ.
\label{eq:t0-mapping}
}
The bond configuration $\bm b=(b_j)_{j\geq1}$ is thereby mapped to a three-state configuration
\eq{
\bm c=(c_j)_{j\geq1},\qquad c_j\in\{\circ,\bullet_1,\bullet_2\}.
}
In these variables, the dynamics takes the simple form
\eq{
\circ\bullet_\alpha\longleftrightarrow\bullet_\alpha\circ,\qquad \alpha=1,2,
}
and is generated by
\eq{
H_{t0}=-\frac12\sum_{j=1}^{\infty}\sum_{\alpha=1,2} \left( \ket{\circ\bullet_\alpha}\bra{\bullet_\alpha\circ} + \ket{\bullet_\alpha\circ}\bra{\circ\bullet_\alpha} \right)_{j,j+1}.
\label{eq:t0-Hamiltonian}
}
Because successive objects are constructed from disjoint sets of
independent bond variables, the infinite-temperature state is mapped
to the product state
\eq{
\rho_{t0} = \bigotimes_{j\geq1} \left[ \frac14\ket{\circ}\bra{\circ} + \frac12\ket{\bullet_1}\bra{\bullet_1} + \frac14\ket{\bullet_2}\bra{\bullet_2} \right]_j.
\label{eq:t0-infinite-temperature-state}
}

The three-state Hamiltonian in Eq.~\eqref{eq:t0-Hamiltonian} is the
Maassarani--Mathieu chain, also known as the $SU(3)$ XX chain~\cite{Massarani_1998}.
Under a Jordan--Wigner transformation, it becomes the $t$--$0$ model.
In this representation, $\circ$ is an empty state, while $\bullet_\alpha$ is an
occupied state carrying the internal label $\alpha=1,2$.
We refer to the occupation degree of freedom (empty or occupied) and the internal label (1 or 2) 
as the $t$--$0$ charge and $t$--$0$ spin, respectively.
The latter terminology distinguishes the internal label from the physical spins of the folded XXZ chain.

A crucial property of the $t$--$0$ model is the exact spin--charge separation~\cite{Ogata_1990, Izergin_1998, Gamayun_2023,Gamayun_2024}.
To make this structure explicit, for each $t$--$0$ configuration $\bm c$, we define the $t$--$0$ charge configuration
\eq{
\bm\eta_c :=(\eta_{c,j})_{j\geq1},\qquad
\eta_{c,j} :=
\begin{cases}
\circ,&c_j=\circ,\\
\bullet,&c_j\in\{\bullet_1,\bullet_2\}.
\end{cases}
}
On this charge space, we define the local occupation operator by
\eq{
n_j:=\ket{\bullet}\bra{\bullet}_j.
\label{eq:def-n}
}
Let $x_1<x_2<\cdots$ denote the positions of the occupied states, $\eta_{c,x_a}=\bullet$.
The corresponding $t$--$0$ spin configuration is defined by
\eq{
\bm\eta_s :=(\eta_{s,1},\eta_{s,2},\ldots),\qquad c_{x_a}=\bullet_{\eta_{s,a}},\qquad \eta_{s,a}\in\{1,2\}.
}
Accordingly, a basis state of the $t$--$0$ model can be represented as
\eq{
\ket{\bm c} = \ket{\bm\eta_c} \otimes\ket{\bm\eta_s},
}
where the second factor is defined on the squeezed lattice formed by the occupied states.

Since the $t$--$0$ dynamics only exchanges $\circ$ with a neighboring
$\bullet_\alpha$, the ordered spin configuration $\bm\eta_s$ is conserved,
while the charge configuration $\bm\eta_c$ evolves independently of
$\bm\eta_s$.
More explicitly,
\eq{
e^{-iH_{t0}t}\ket{\bm\eta_c}\otimes\ket{\bm\eta_s} = \left(e^{-iH_{\xx}t}\ket{\bm\eta_c}\right)\otimes\ket{\bm\eta_s},
}
where
\eq{
H_{\xx} = -\frac12\sum_{j=1}^{\infty} \left( \ket{\circ\bullet}\bra{\bullet\circ} + \ket{\bullet\circ}\bra{\circ\bullet} \right)_{j,j+1}.
\label{eq:charge-XX-Hamiltonian}
}
Thus, the $t$--$0$ spin configuration is static on the squeezed lattice,
whereas the $t$--$0$ charge dynamics is governed by the XX chain and
hence reduces to a free-fermion problem.

In this spin--charge representation, the infinite-temperature state also factorizes as
\eq{
\rho_{t0}=\rho_c\otimes\rho_s,
\label{eq:t0-state-factorization}
}
where
\eq{
\rho_c=
\bigotimes_{j\geq1}
\left[\frac14\ket{\circ}\bra{\circ}+\frac34\ket{\bullet}\bra{\bullet}\right]_j,
\qquad
\rho_s =
\bigotimes_{a\geq1}
\left[\frac23\ket{1}\bra{1}+\frac13\ket{2}\bra{2}\right]_a.
\label{eq:t0-charge-spin-states}
}
Thus, the $t$--$0$ charge configuration and the $t$--$0$ spin configuration
are statistically independent in the infinite-temperature state.

This spin--charge separation is central to the calculation of the current MGF.
As shown below, the static $t$--$0$ spin configuration gives rise to a
combinatorial contribution, whereas the dynamical part is entirely determined by the XX chain.

\subsection{Representation of \texorpdfstring{$S_R$}{S_R} in the $t$--$0$ model}

We next express $S_R^{(\sigma)}$ in terms of the $t$--$0$ variables.
Since the mapping in Eq.~\eqref{eq:t0-mapping} changes the length of the configuration, 
the number of $t$--$0$ objects associated with the finite bond variables depends on the values of these bond variables.

\begin{figure}[t]
\centering

\begin{minipage}{0.49\linewidth}
\centering
\begin{tikzpicture}[
    x=0.60cm,
    y=0.9cm,
    font=\normalsize,
    bond/.style={draw, minimum height=0.50cm, inner sep=0pt},
    mover/.style={draw, rounded corners=2pt},
    zeroone/.style={draw, rounded corners=2pt},
    zerotwo/.style={draw, rounded corners=2pt},
    cut/.style={densely dashed, thick},
    counted/.style={fill=blue!12},
]

\node at (4.5,3.3) {\textbf{(a)} $N_1$ even};

\draw[decorate,decoration={brace,amplitude=4pt}]
    (0,2.15) -- (8,2.15)
    node[midway,above=5pt] {$[0,R]$};

\node[anchor=east] at (-0.35,1.75) {folded XXZ};
\node[anchor=east] at (-0.35,0.85) {bond};
\node[anchor=east] at (-0.35,0.00) {$t$--$0$};

\foreach \x/\s in {
0/\uparrow,
1/\downarrow,
2/\uparrow,
3/\uparrow,
4/\downarrow,
5/\downarrow,
6/\downarrow,
7/\uparrow,
8/\downarrow,
9/\downarrow
}{
    \node at (\x,1.75) {\large$\boldsymbol{\s}$};
}

\foreach \x/\b in {
0/1,
1/1,
2/0,
3/1,
4/0,
5/0,
6/1,
7/1,
8/0
}{
    \draw[bond] (\x,0.60) rectangle ++(1,0.5);
    \node at (\x+0.5,0.85) {$\b$};
}

\draw[mover,counted]   (0,-0.25) rectangle (2,0.25);
\node at (1,0) {$\circ$};

\draw[zeroone,counted] (2,-0.25) rectangle (3,0.25);
\node at (2.5,0) {$\bullet_1$};

\draw[zerotwo,counted] (3,-0.25) rectangle (5,0.25);
\node at (4,0) {$\bullet_2$};

\draw[zeroone,counted] (5,-0.25) rectangle (6,0.25);
\node at (5.5,0) {$\bullet_1$};

\draw[mover,counted]   (6,-0.25) rectangle (8,0.25);
\node at (7,0) {$\circ$};

\draw[zeroone] (8,-0.25) rectangle (9,0.25);
\node at (8.5,0) {$\bullet_1$};

\draw[cut] (8,-0.55) -- (8,2.35);

\end{tikzpicture}
\end{minipage}
\hfill
\begin{minipage}{0.48\linewidth}
\centering
\begin{tikzpicture}[
    x=0.60cm,
    y=0.9cm,
    font=\normalsize,
    bond/.style={draw, minimum height=0.50cm, inner sep=0pt},
    mover/.style={draw, rounded corners=2pt},
    zeroone/.style={draw, rounded corners=2pt},
    zerotwo/.style={draw, rounded corners=2pt},
    cut/.style={densely dashed, thick},
    counted/.style={fill=blue!12},
]

\node at (4.5,3.3) {\textbf{(b)} $N_1$ odd};

\draw[decorate,decoration={brace,amplitude=4pt}]
    (0,2.15) -- (8,2.15)
    node[midway,above=5pt] {$[0,R]$};

\foreach \x/\s in {
0/\uparrow,
1/\downarrow,
2/\uparrow,
3/\uparrow,
4/\downarrow,
5/\downarrow,
6/\uparrow,
7/\downarrow,
8/\uparrow,
9/\uparrow
}{
    \node at (\x,1.75) {\large$\boldsymbol{\s}$};
}

\foreach \x/\b in {
0/1,
1/1,
2/0,
3/1,
4/0,
5/1,
6/1,
7/1,
8/0
}{
    \draw[bond] (\x,0.60) rectangle ++(1,0.5);
    \node at (\x+0.5,0.85) {$\b$};
}

\draw[mover,counted]   (0,-0.25) rectangle (2,0.25);
\node at (1,0) {$\circ$};

\draw[zeroone,counted] (2,-0.25) rectangle (3,0.25);
\node at (2.5,0) {$\bullet_1$};

\draw[zerotwo,counted] (3,-0.25) rectangle (5,0.25);
\node at (4,0) {$\bullet_2$};

\draw[mover,counted]   (5,-0.25) rectangle (7,0.25);
\node at (6,0) {$\circ$};

\draw[zerotwo] (7,-0.25) rectangle (9,0.25);
\node at (8,0) {$\bullet_2$};

\draw[cut] (8,-0.55) -- (8,2.35);

\end{tikzpicture}
\end{minipage}

\caption{
Schematic illustration of the mapping from the folded XXZ spin configuration to the bond configuration and then to the $t$--$0$ configuration.
For even $N_1$, the endpoint $R\in 2\mathbb N$ coincides with a boundary between two $t$--$0$ objects, whereas for odd $N_1$, it lies inside a $t$--$0$ object.
We denote by $L$, $N$, and $N_1$ the numbers of $t$--$0$ objects, $\bullet$-objects, and $\bullet_1$-objects, respectively,
that are fully contained in $(b_1,\ldots,b_R)$.
These complete objects are shaded in blue.
A $t$--$0$ object straddling the endpoint $R$ is not counted.
For panels (a) and (b), the corresponding sectors are $(L,N,N_1)=(5,3,2)$ and $(L,N,N_1)=(4,2,1)$, respectively.
}
\label{fig:mapping}
\end{figure}
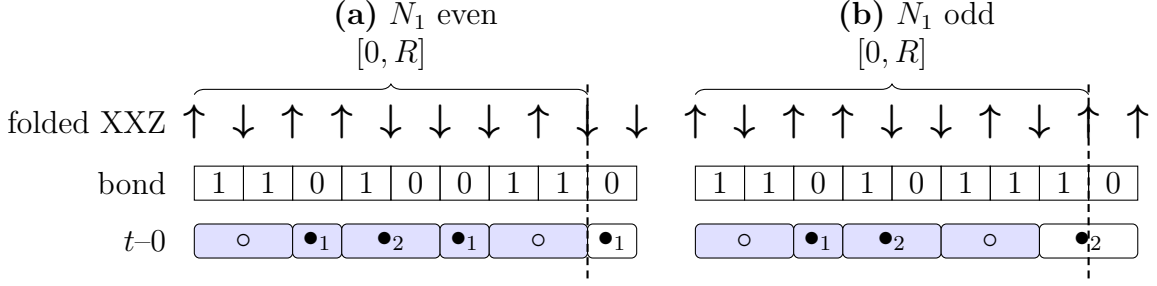

Let $L$ denote the number of complete $t$--$0$ objects contained in
$(b_1,\ldots,b_R)$, and let $N$ and $N_1$ denote, respectively,
the numbers of occupied objects and occupied objects with $t$--$0$ spin label $1$
among them.
Thus, the numbers of $\circ$-objects and $\bullet_2$-objects are $L-N$ and $N-N_1$, respectively.
If the endpoint $R\in 2\mathbb{N}$ lies inside a $\circ$-object or a $\bullet_2$-object,
that object is not counted, as illustrated in Fig.~\ref{fig:mapping}.

In the spin--charge representation introduced above,
the $N$ occupied objects in the interval carry the first $N$ entries,
\eq{
(\eta_{s,1},\ldots,\eta_{s,N}),
}
of the $t$--$0$ spin configuration $\bm\eta_s$.
As we show below, for fixed $\sigma$ and $N$,
$S_R^{(\sigma)}$ is determined entirely by these spin labels
and is independent of the charge configuration $\bm\eta_c$.

We first consider $N_1\in2\mathbb Z_{\geq 0}$.
In this case, the $t$--$0$ objects contained in the interval account for all $R$ bond variables; see Fig.~\ref{fig:mapping} (a), and therefore
\eq{
2(L-N)+N_1+2(N-N_1)=R.
\label{eq:length-constraint-even}
}
No $t$--$0$ object straddles the endpoint.
Using Eq.~\eqref{eq:SR-bond}, we obtain
\eq{
S_R^{(\sigma)} = \frac{\sigma}{2}\sum_{j=1}^{L}\prod_{k<j}(-1)^{\ind{c_k=\bullet_2}}\ind{c_j=\bullet_1} + \frac{\sigma}{2}\prod_{k=1}^{L}(-1)^{\ind{c_k=\bullet_2}}.
}
Here, $\chi[\cdot]$ is the indicator function.
Since the $\circ$-objects do not contribute to either term, this expression
can be written entirely in terms of the $t$--$0$ spin configuration as
\eq{
S_R^{(\sigma)}
=
\frac{\sigma}{2}\sum_{j=1}^{N}
\prod_{k<j}(-1)^{\ind{\eta_{s,k}=2}}\ind{\eta_{s,j}=1}
+\frac{\sigma}{2}\prod_{k=1}^{N}(-1)^{\ind{\eta_{s,k}=2}}.
\label{eq:SR-t0-even}
}

We next consider $N_1\in2\mathbb Z_{\geq 0} + 1$.
In this case,
\eq{
2(L-N)+N_1+2(N-N_1)=R-1.
\label{eq:length-constraint-odd}
}
Thus, the $t$--$0$ variables $c_1,\ldots,c_L$ account for the first $R-1$ bond variables, 
while the endpoint lies inside the next object $c_{L+1}$; see Fig.~\ref{fig:mapping} (b).
Since a $\bullet_1$-object consists of a single bond, the object straddling the endpoint must be either $\circ$ or $\bullet_2$, and hence
\eq{
c_{L+1}\in\{\circ,\bullet_2\}.
\label{eq:endpoint-odd}
}
Proceeding as in the even case, we find
\eq{
S_R^{(\sigma)} = \frac{\sigma}{2}\sum_{j=1}^{N}\prod_{k<j}(-1)^{\ind{\eta_{s,k}=2}}\ind{\eta_{s,j}=1}.
\label{eq:SR-t0-odd}
}

The two length constraints, Eqs.~\eqref{eq:length-constraint-even} and \eqref{eq:length-constraint-odd}, can be combined into
\eq{
2L-N_1 = R-\frac12\left[1-(-1)^{N_1}\right].
\label{eq:length-constraint}
}
Similarly, Eqs.~\eqref{eq:SR-t0-even} and \eqref{eq:SR-t0-odd} can be written in the unified form
\eq{
S_R^{(\sigma)}
=
\frac{\sigma}{2}\sum_{j=1}^{N}
\prod_{k<j}(-1)^{\ind{\eta_{s,k}=2}}\ind{\eta_{s,j}=1}
+\frac{\sigma}{4}\left[1+(-1)^{N_1}\right](-1)^{N-N_1}.
\label{eq:SR-t0}
}
Thus, for a fixed boundary spin $\sigma$, the physical magnetization $S_R^{(\sigma)}$ 
depends only on the first $N$ entries of the static $t$--$0$ spin configuration $\bm\eta_s$.

\subsection{Physical meaning of $S_R$}
We emphasize that $S_R^{(\sigma)}$ in Eq.~\eqref{eq:SR-t0}
does not correspond to the usual spin magnetization in the $t$--$0$ model.
Each spin label $1$ contributes with a sign that is reversed
whenever a spin label $2$ is encountered.
The sum therefore counts the lengths of consecutive runs of label $1$
with alternating signs.
For example, for the spin configuration
\eq{
\bm\eta_s=(1,1,2,1,1,2,1,2,\cdots),
}
Eq.~\eqref{eq:SR-t0} takes the form
\eq{
S_R^{(\sigma)}=\frac{\sigma}{2}\left[2-2+1-\cdots\right].
}
This observable differs from the usual $t$--$0$ spin magnetization,
which assigns a fixed value of $+1/2$ or $-1/2$ to each spin label.

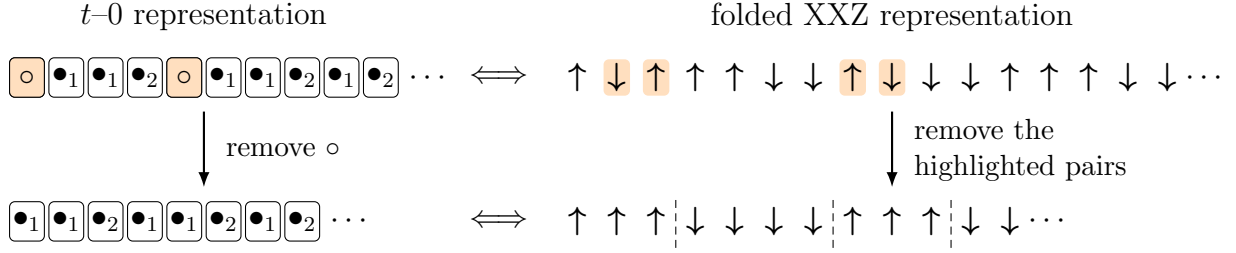
\begin{figure}[t]
\centering
\begin{tikzpicture}[
    x=0.52cm,
    y=1cm,
    font=\normalsize,
    object/.style={
        draw,
        rounded corners=2pt,
        minimum width=0.46cm,
        minimum height=0.52cm,
        inner sep=1pt
    },
    removed/.style={fill=orange!25},
    spin/.style={
        minimum width=0.35cm,
        minimum height=0.48cm,
        inner sep=0pt
    },
    arrow/.style={-{Latex[length=2mm]}, thick},
]

\node at (4.5,3.0) {$t$--$0$ representation};
\node at (22,3.0) {folded XXZ representation};

\foreach \x/\s in {
    0/\circ,
    1/\bullet_1,
    2/\bullet_1,
    3/\bullet_2,
    4/\circ,
    5/\bullet_1,
    6/\bullet_1,
    7/\bullet_2,
    8/\bullet_1,
    9/\bullet_2
}{
    \node[object] at (\x,2.2) {$\s$};
}
\node[object,removed] at (0,2.2) {$\circ$};
\node[object,removed] at (4,2.2) {$\circ$};
\node at (10.25,2.2) {$\cdots$};

\foreach \x/\s in {
    14/\uparrow,
    15/\downarrow,
    16/\uparrow,
    17/\uparrow,
    18/\uparrow,
    19/\downarrow,
    20/\downarrow,
    21/\uparrow,
    22/\downarrow,
    23/\downarrow,
    24/\downarrow,
    25/\uparrow,
    26/\uparrow,
    27/\uparrow,
    28/\downarrow,
    29/\downarrow
}{
    \node[spin] at (\x,2.2) {$\boldsymbol{\s}$};
}
\foreach \x/\s in {
    15/\downarrow,
    16/\uparrow,
    21/\uparrow,
    22/\downarrow
}{
    \node[spin,removed,rounded corners=2pt]
        at (\x,2.2) {$\boldsymbol{\s}$};
}
\node at (30,2.2) {$\cdots$};

\node at (12,2.2) {$\Longleftrightarrow$};

\draw[arrow] (4.5,1.75) -- (4.5,0.75)
    node[midway,right=4pt,align=left,font=\small]
    {remove $\circ$};

\draw[arrow] (22,1.75) -- (22,0.75)
    node[midway,right=4pt,align=left,font=\small]
    {remove the\\highlighted pairs};

\foreach \x/\s in {
    0/\bullet_1,
    1/\bullet_1,
    2/\bullet_2,
    3/\bullet_1,
    4/\bullet_1,
    5/\bullet_2,
    6/\bullet_1,
    7/\bullet_2
}{
    \node[object] at (\x,0.3) {$\s$};
}
\node at (8.25,0.3) {$\cdots$};

\foreach \x/\s in {
    14/\uparrow,
    15/\uparrow,
    16/\uparrow,
    17/\downarrow,
    18/\downarrow,
    19/\downarrow,
    20/\downarrow,
    21/\uparrow,
    22/\uparrow,
    23/\uparrow,
    24/\downarrow,
    25/\downarrow
}{
    \node[spin] at (\x,0.3) {$\boldsymbol{\s}$};
}
\node at (26,0.3) {$\cdots$};

\foreach \x in {16.5,20.5,23.5}{
    \draw[densely dashed] (\x,-0.05) -- (\x,0.65);
}

\node at (12,0.3) {$\Longleftrightarrow$};

\end{tikzpicture}

\caption{
Relation between the $t$--$0$ spin configuration and the magnetization
domains in the folded XXZ chain.
The upper row shows a $t$--$0$ configuration and its corresponding
folded spin configuration, with the leftmost spin fixed to $\uparrow$.
Removing the $\circ$-objects eliminates the highlighted
oppositely oriented spin pairs, leaving the total magnetization unchanged.
The lower row shows the remaining occupied objects and the
corresponding folded configuration, which consists of magnetization domains without isolated spins.
The dashed lines mark the positions of the domain walls,
each represented by a $t$--$0$ spin label $2$.
Cancelling an oppositely oriented spin pair across each wall
leaves successive groups of two up spins, two down spins, and one up spin, and so on.
These correspond to the consecutive runs of label $1$ in
$\bm\eta_s=(1,1,2,1,1,2,1,2,\cdots)$, giving
$S_R^{(+)}=\frac12[2-2+1-\cdots]$.
}
\label{fig:spin-domain-interpretation}
\end{figure}

This alternating sum also has an interpretation in terms of spin domains in the folded XXZ chain, as illustrated
in Fig.~\ref{fig:spin-domain-interpretation}.
A $\circ$-object corresponds to the local bond configuration $11$, 
or a local spin configuration $\uparrow\downarrow\uparrow$ (or its spin-reversed counterpart).
Removing this object eliminates an oppositely oriented spin pair and leaves the total magnetization unchanged.
Thus, removing all $\circ$-objects gives a representative folded
configuration determined by the $t$--$0$ spin configuration alone, with the same value of $S_R^{(\sigma)}$.
In this configuration, neither $\uparrow\downarrow\uparrow$ nor $\downarrow\uparrow\downarrow$ occurs:
the isolated spins represented by the $\circ$-objects have been removed.
The remaining configuration therefore consists of magnetization domains separated by domain walls.

Each spin label $2$ marks a domain wall, across which the magnetization reverses its sign.
After cancelling the oppositely oriented spin pairs at the domain boundaries, the remaining spins are counted by the
consecutive runs of label $1$.
Their contributions therefore alternate in sign, giving the alternating sum in Eq.~\eqref{eq:SR-t0}.

\section{Exact current MGF}
\label{sec:exact-mgf}

\subsection{Sector decomposition}
\label{subsec:decomp}

As shown in the previous section, the representation of $S_R^{(\sigma)}$ in the $t$--$0$ model
is naturally organized by the numbers of $t$--$0$ objects associated with the bond configuration $(b_1,\ldots,b_R)$.
We label the initial and final sectors by
\eq{
\omega=(L,N,N_1), \qquad \omega'=(L',N',N_1').
}
Here, $L$ and $L'$ are the numbers of complete $t$--$0$ objects contained within the first $R$ bond variables at the initial and final times, respectively.
Among them, $N$ and $N'$ are the numbers of occupied objects, and $N_1$ and $N_1'$
are the numbers of occupied objects carrying the $t$--$0$ spin label $1$.
The admissible sectors satisfy
\eq{
L\in\mathbb Z_{\geq0}, \qquad 0\leq N_1\leq N\leq L,
}
together with Eq.~\eqref{eq:length-constraint}.
We denote the set of admissible sectors by $\Omega$.
Below, we rewrite the MGF~\eqref{eq:half-infinite-mgf} in the $t$--$0$ representation
by decomposing the initial and final states into sectors labeled by $\omega$ and $\omega'$, respectively.

To express the MGF in the $t$--$0$ representation, we first rewrite it in the bond representation.
In the bond model, the Hamiltonian, the initial state, and the magnetization in the interval $[0,R]$ are
given by Eqs.~\eqref{eq:bond-Hamiltonian}, \eqref{eq:bond-infinite-temperature-state}, and \eqref{eq:SR-bond}, respectively.
The current MGF can therefore be written as
\eq{
\left\langle e^{\lambda Q_t}\right\rangle_R =
\frac12\sum_{\sigma=\pm1}
\Tr_b\left[
e^{\lambda S_R^{(\sigma)}}e^{-iH_bt} e^{-\lambda S_R^{(\sigma)}}\rho_b e^{iH_bt}
\right],
\label{eq:mgf-bond}
}
where $\Tr_b$ denotes the trace in the bond representation.
Note that under the bond mapping \eqref{eq:bond-map}, 
the boundary spin $\sigma$ is independent of the bond configuration and 
takes the values $\pm1$ with equal probability,
since the infinite-temperature measure is uniform in the spin variables.

Let $P_\omega$ denote the projector onto the sector $\omega$.
In the bond representation, it is given by
\eq{
P_\omega =  \sum_{b_1,\ldots,b_R \in\{0,1\}} \chi_{\omega}(b_1,\ldots,b_R)\ket{b_1,\ldots,b_R}\bra{b_1,\ldots,b_R},
}
where $\chi_\omega(b_1,\ldots,b_R)$ is equal to $1$ if the bond configuration
$(b_1,\ldots,b_R)$ belongs to the sector $\omega$, and is equal to $0$ otherwise.
Here and below, identity operators on the remaining semi-infinite degrees of freedom are left implicit.
Let $I_b$ denote the identity operator on the bond Hilbert space.
Inserting the sector identity $\sum_{\omega \in \Omega} P_{\omega}=I_b$ into Eq.~\eqref{eq:mgf-bond}, we obtain
\eq{
\braket{e^{\lambda Q_t}}_R = \sum_{\omega,\omega'\in\Omega} G(\lambda\mid\omega';\omega),
\label{eq:mgf-sector-decomposition}
}
where
\eq{
G(\lambda\mid\omega';\omega)
=
\frac12\sum_{\sigma=\pm1}
\Tr_b\left[
e^{\lambda S_R^{(\sigma)}}P_{\omega'}e^{-iH_bt}
e^{-\lambda S_R^{(\sigma)}}P_{\omega}\rho_b e^{iH_bt}
\right].
\label{eq:def-G}
}

We now rewrite Eq.~\eqref{eq:def-G} in the spin--charge representation.
To this end, we first express the sector projector $P_\omega$ in terms of the charge and spin variables.
Let
\eq{
P_N^L
:=
\sum_{\substack{
\eta_{c,1},\ldots,\eta_{c,L}\in \{\circ,\bullet\}\\
\sum_{j=1}^{L}\ind{\eta_{c,j}=\bullet}=N
}}
\ket{\eta_{c,1},\ldots,\eta_{c,L}}
\bra{\eta_{c,1},\ldots,\eta_{c,L}}
\label{eq:charge-sector-projector}
}
be the projector onto charge configurations containing exactly $N$
occupied states among the first $L$ sites.
Similarly, we define
\eq{
P_{N_1}^{N} :=
\sum_{\substack{
\eta_{s,1},\ldots,\eta_{s,N} \in \{1,2\}\\
\sum_{j=1}^{N}\ind{\eta_{s,j}=1}=N_1
}}
\ket{\eta_{s,1},\ldots,\eta_{s,N}} \bra{\eta_{s,1},\ldots,\eta_{s,N}}.
\label{eq:spin-sector-projector}
}

For even $N_1$, no $t$--$0$ object straddles the endpoint, and hence
\eq{
P_\omega = P_N^L\otimes P_{N_1}^{N}.
}
For odd $N_1$, $c_{L+1}$ must be either $\circ$ or $\bullet_2$.
In the spin--charge representation, this condition is imposed by
\eq{
E_{L+1,N+1}
:=
\ket{\circ}\bra{\circ}_{L+1}\otimes I_s
+\ket{\bullet}\bra{\bullet}_{L+1}\otimes\ket{2}\bra{2}_{N+1},
\label{eq:endpoint-spin-charge-projector}
}
where $I_s$ denotes the identity operator on the $t$--$0$ spin space.
Introducing
\eq{
\epsilon := \frac{1-(-1)^{N_1}}{2},
}
the sector projector can therefore be written as
\eq{
P_\omega = \left(P_N^L\otimes P_{N_1}^{N}\right) E_{L+1,N+1}^{\epsilon }.
\label{eq:sector-projector}
}

We next incorporate the magnetization weight into the spin projector.
For $\bm{\eta} \in \{1,2\}^N $, define
\eq{
F_{N,N_1}(\bm{\eta}) := \frac12\sum_{j=1}^{N}
\prod_{k<j}(-1)^{\ind{\eta_k=2}}\ind{\eta_j=1} +\frac14\left[1+(-1)^{N_1}\right](-1)^{N-N_1}.
\label{eq:def-F-sector}
}
Equation~\eqref{eq:SR-t0} then gives
\eq{
S_R^{(\sigma)} = \sigma F_{N,N_1}(\eta_{s,1},\ldots,\eta_{s,N})
}
for a state $\ket{\bm\eta_c}\otimes\ket{\bm\eta_s}$ in the sector
$\omega=(L,N,N_1)$.
We therefore introduce the tilted spin projector
\eq{
P_{N_1}^{N}(\mu) :=
\sum_{\substack{
\eta_{s,1},\ldots,\eta_{s,N}\in\{1,2\}\\
\sum_{j=1}^{N}\ind{\eta_{s,j}=1}=N_1
}}
e^{\mu F_{N,N_1}(\eta_{s,1},\ldots,\eta_{s,N})}
\ket{\eta_{s,1},\ldots,\eta_{s,N}} \bra{\eta_{s,1},\ldots,\eta_{s,N}}.
\label{eq:tilted-spin-projector}
}
The corresponding tilted sector projector is defined by
\eq{
P_\omega^{(\mu)} := \left( P_N^L\otimes P_{N_1}^{N}(\mu) \right) E_{L+1,N+1}^{\epsilon }.
\label{eq:tilted-sector-projector}
}
By construction,
\eq{
e^{\lambda S_R^{(\sigma)}}P_\omega = P_\omega^{(\sigma\lambda)}.
\label{eq:tilted-sector-identity}
}

Using Eq.~\eqref{eq:tilted-sector-identity}, 
together with the spin--charge separation of the dynamics and the initial state,
Eqs.~\eqref{eq:charge-XX-Hamiltonian} and \eqref{eq:t0-state-factorization}, we obtain
\eq{
G(\lambda\mid\omega';\omega) =
\frac12\sum_{\sigma=\pm1}
\Tr_{t0}\left[
P_{\omega'}^{(\sigma\lambda)}
\left(e^{-iH_{\xx}t}\otimes I_s\right)
P_{\omega}^{(-\sigma\lambda)}
(\rho_c\otimes\rho_s)
\left(e^{iH_{\xx}t}\otimes I_s\right)
\right],
\label{eq:G-spin-charge}
}
where $\Tr_{t0}$ denotes the trace in the $t$--$0$ representation.

Equation~\eqref{eq:G-spin-charge} implies that the sector contribution is
symmetric under the exchange of the initial and final sectors.
Indeed, in the spin--charge basis, $H_{\xx}$ is real and symmetric, whereas
$P_{\omega}^{(\mu)}$, $\rho_c$, and $\rho_s$ are real and diagonal.
Moreover, $\rho_c$ commutes with $H_{\xx}$.
Taking the transpose of the trace in Eq.~\eqref{eq:G-spin-charge}, using
cyclicity of the trace, and changing $\sigma\to-\sigma$, we obtain
\eq{
G(\lambda\mid\omega';\omega)
=
G(\lambda\mid\omega;\omega').
\label{eq:G-time-reversal}
}
Consequently, the contributions with $n:=N'-N<0$ are identical to those
with $n>0$ after exchanging $\omega$ and $\omega'$.

When $N'=N$, every sector contribution with nonzero transition weight is diagonal.
Indeed, since a $t$--$0$ spin configuration is conserved under the dynamics, the equality $N'=N$ implies $N_1'=N_1$ whenever
$G(\lambda\mid\omega';\omega)\neq0$.
The constraint~\eqref{eq:length-constraint} then gives $L'=L$.
Hence,
\eq{
G(\lambda\mid\omega';\omega)\neq0,\quad N'=N
\qquad\Longrightarrow\qquad
\omega'=\omega.
\label{eq:diagonal-sector}
}
The moment-generating function can therefore be rewritten as
\eq{
\braket{e^{\lambda Q_t}}_R
=
\sum_{\omega\in\Omega}G(\lambda\mid\omega;\omega)
+
2\sum_{\substack{\omega,\omega'\in\Omega\\ N'-N>0}}
G(\lambda\mid\omega';\omega).
\label{eq:MGF-time-reversal}
}

\subsection{Spin--charge factorization of the MGF}
Here, we show that $G(\lambda\mid\omega';\omega)$ decomposes into
products of two factors that depend only on the $t$--$0$ charge and
$t$--$0$ spin degrees of freedom, respectively.

To illustrate this spin--charge factorization, consider first the case $N_1,N_1'\in2\mathbb Z_{\geq0}$.
In this case, no endpoint constraint is present in the tilted projector \eqref{eq:tilted-sector-projector} and
\eq{
P_{\omega}^{(\mu)} = P_N^L\otimes P_{N_1}^{N}(\mu), \qquad P_{\omega'}^{(\mu)} = P_{N'}^{L'}\otimes P_{N_1'}^{N'}(\mu).
}
Equation~\eqref{eq:G-spin-charge} therefore factorizes as
\eqs{
G(\lambda\mid\omega';\omega) =
\Tr_c\left[
P_{N'}^{L'}e^{-iH_{\xx}t}P_N^L\rho_c e^{iH_{\xx}t}
\right]
\frac12\sum_{\sigma=\pm1}
\Tr_s\left[
P_{N_1'}^{N'}(\sigma\lambda)P_{N_1}^{N}(-\sigma\lambda)\rho_s
\right],
\label{eq:G-factorization-even}
}
where $\Tr_c$ and $\Tr_s$ denote the traces over the $t$--$0$ charge and spin spaces, respectively.

The two factors in Eq.~\eqref{eq:G-factorization-even} have completely different origins.
The first factor is the transition weight from $(L,N)$ to $(L',N')$
under the XX-chain dynamics and can therefore be treated as a free-fermion problem.
By contrast, the $t$--$0$ spin configuration is static, so that the
second factor contains no time evolution and reduces to a combinatorial average over the spin configurations.
Equivalently, in the folded XXZ representation, the first factor describes the dynamics of the isolated spins, 
which correspond to the $\circ$-objects in the $t$--$0$ model.
Removing these objects leaves a static configuration of magnetization domains, as illustrated in Fig.~\ref{fig:spin-domain-interpretation}.
The second factor is the combinatorial average over these domain configurations.

When either $N_1$ or $N_1'$ is odd, the corresponding sector projector contains the local endpoint constraint introduced in
Eq.~\eqref{eq:endpoint-spin-charge-projector}.
Expanding the endpoint projectors expresses the sector contribution as a
finite sum of products of XX-chain charge traces and static spin traces.

As shown in Appendices~\ref{app:sector-spin-traces} and \ref{app:sector-combinatorics}, the static spin traces can be evaluated combinatorially.
For $N'>N$, we define
\eq{
n:=N'-N>0,\qquad k:=N_1'-N_1.
}
Because the spin configuration is conserved, the additional $n$ spin
labels contain exactly $k$ labels equal to $1$, and hence $0\leq k\leq n$.
We also define initial and final parities
\eq{
\epsilon_{\rm i} :=
\frac{1-(-1)^{N_1}}{2},
\qquad
\epsilon_{\rm f} :=
\frac{1-(-1)^{N_1'}}{2} =
(\epsilon_{\rm i}+k)\bmod 2.
}
With these definitions, $G(\lambda\mid\omega';\omega)$ for $n>0$ is given by
\eq{
&G(\lambda\mid\omega';\omega) = \mathcal{P}_c(L,N)\mathcal{P}_s(N,N_1)B_{n,k}
\\
&
\times
\begin{cases}
p_t^{(\varnothing,\epsilon_{\rm f})}(L',N';L,N)
\mathcal G_{\lambda}^{(0)}(n,k),
& \epsilon_{\rm i}=0,
\\
p_t^{(\circ,\epsilon_{\rm f})}(L',N';L,N)
\mathcal G_{\lambda}^{(1)}(n,k)
+\dfrac{n-k}{n} p_t^{(\bullet,\epsilon_{\rm f})}(L',N';L,N) \mathcal G_{\lambda}^{(1)}(n-1,k),
& \epsilon_{\rm i}=1.
\end{cases}
\label{eq:G-sector-final}
}
Here, $\mathcal P_c(L,N)$ and $\mathcal P_s(N,N_1)$ are the combinatorial weights
\eq{
\mathcal P_c(L,N) :=
\binom LN\left(\frac34\right)^N\left(\frac14\right)^{L-N},
\qquad
\mathcal P_s(N,N_1) :=
\binom{N}{N_1}\left(\frac23\right)^{N_1} \left(\frac13\right)^{N-N_1}.
\label{eq:initial-weights}
}
The factor $B_{n,k}$ is defined in Eq.~\eqref{eq:binom-B}, whereas
$\mathcal G_\lambda^{(\epsilon)}(n,k)$ is defined in Eq.~\eqref{eq:main-Gcal}.

The charge factor $p_t^{(r,\epsilon)}(L',N';L,N)$, where
$r\in\{\varnothing,\circ,\bullet\}$ specifies the initial endpoint condition and
$\epsilon\in\{0,1\}$ specifies whether the final endpoint constraint is
present, is defined by
\eq{
p_t^{(r,\epsilon)}(L',N';L,N)
:=
\frac{1}{\mathcal P_c(L,N)}
\Tr_c\left[
E_{L'+1}^{\epsilon}P_{N'}^{L'}e^{-iH_{\xx}t}
P_N^L C_{L+1}^{(r)}\rho_c e^{iH_{\xx}t}
\right].
\label{eq:charge-factor-main}
}
Here,
\eq{
C_j^{(\varnothing)}:=I_c,\qquad
C_j^{(\circ)}:=\ket{\circ}\bra{\circ}_j,\qquad
C_j^{(\bullet)}:=\ket{\bullet}\bra{\bullet}_j,
\label{eq:def-C-pro}
}
and
\eq{
E_j:=\ket{\circ}\bra{\circ}_j+\frac13\ket{\bullet}\bra{\bullet}_j,
\label{eq:def-E-pro}
}
where $I_c$ denotes the identity operator on the $t$--$0$ charge space.
For a diagonal sector, the opposite tilts in the two spin projectors cancel,
\eq{
P_{N_1}^{N}(\sigma\lambda)P_{N_1}^{N}(-\sigma\lambda) = P_{N_1}^{N},
}
and hence
\eq{
G(\lambda\mid\omega;\omega) = G(0\mid\omega;\omega).
\label{eq:G-diagonal-lambda-independent}
}
Combining this equation with Eq.~\eqref{eq:MGF-time-reversal} and using
$\braket{e^{\lambda Q_t}}_R|_{\lambda=0}=1$, we obtain
\eq{
\braket{e^{\lambda Q_t}}_R = 1+
2\sum_{\substack{\omega,\omega'\in\Omega\\ N'-N>0}}
\left[
G(\lambda\mid\omega';\omega)-G(0\mid\omega';\omega)
\right].
\label{eq:MGF-positive-n}
}

\subsection{Infinite-system limit}

We now take the infinite-system limit $R\to\infty$ in
Eqs.~\eqref{eq:G-sector-final} and \eqref{eq:MGF-positive-n}.
By Eq.~\eqref{eq:mgf-bulk-limit}, this yields the infinite-system MGF
$\braket{e^{\lambda Q_t}}$.
We take this limit with
\eq{
n=N'-N>0,\qquad k=N_1'-N_1,
}
and the initial parity
\eq{
\epsilon_{\rm i}=\frac{1-(-1)^{N_1}}{2}
}
fixed.
The differences between the initial and final charge sectors then remain finite:
\eq{
L'-L=x_{\epsilon_{\rm i}}(k),\qquad N'-N=n,
}
where $x_\epsilon(k)$ is defined in Eq.~\eqref{eq:def-x}.
We note 
\eq{
x_{\epsilon_{\rm i}}(k) \geq 0
} 
for $n\geq k \geq 0$ by the length constraint~\eqref{eq:length-constraint}.

The weights $\mathcal P_c(L,N)\mathcal P_s(N,N_1)$ concentrate on sectors satisfying
\eq{
\frac{N}{L}\longrightarrow\frac34,
\qquad
\frac{N_1}{N}\longrightarrow\frac23
}
as $R\to\infty$.
Note that the length constraint~\eqref{eq:length-constraint} also implies $L\to\infty$ in this limit.
As shown in Appendix~\ref{app:bulk-transition}, $p_t^{(r,\epsilon)}(L+x,N+n;L,N)$ has the infinite-system limit
\eq{
\lim_{\substack{L,N\to\infty\\ N/L\to3/4}} p_t^{(r,\epsilon)}(L+x,N+n;L,N) = p_t^{(r,\epsilon)}(x,n),
\label{eq:bulk-limit-p}
}
where
\eq{
p_t^{(r,\epsilon)}(x,n) := \int_{-\pi}^{\pi}\frac{d\theta}{2\pi} e^{-in\theta}F_t^{(r,\epsilon)}(\theta;x)
\label{eq:bulk-p-fourier}
}
and
\eq{
F_t^{(r,\epsilon)}(\theta;x) :=
\Tr_c\left[
E_{x+1}^{\epsilon}e^{i\theta\mathcal N_x}
e^{-iH_{\xx}t}
e^{-i\theta\mathcal N_0}C_1^{(r)}
\rho_c e^{iH_{\xx}t}
\right].
\label{eq:def-F}
}
In these expressions, $H_{\xx}$ and $\rho_c$ are the extensions
of Eqs.~\eqref{eq:charge-XX-Hamiltonian} and
\eqref{eq:t0-charge-spin-states}, respectively, to the full lattice
$\mathbb Z$, and $\mathcal N_x:=\sum_{j\leq x}n_j$.

In the infinite-system limit, for fixed $n$, $k$, and $\epsilon_{\rm i}$, 
the factors $p_t^{(r,\epsilon_{\rm f})}(x_{\epsilon_{\rm i}}(k),n)$ and $\mathcal G_\lambda^{(\epsilon_{\rm i})}(n,k)$
are independent of $L$, $N$, and $N_1$.
Thus, apart from the initial parity $\epsilon_{\rm i}$,
the dependence of Eq.~\eqref{eq:G-sector-final} on the initial sector is contained entirely in
$\mathcal P_c(L,N)\mathcal P_s(N,N_1)$.
For each $\epsilon\in\{0,1\}$, the total weight of the initial sectors
with $\epsilon_{\rm i}=\epsilon$ satisfies
\eq{
\lim_{R\to\infty}
\sum_{\substack{\omega\in\Omega\\ \epsilon_{\rm i}=\epsilon}}
\mathcal P_c(L,N)\mathcal P_s(N,N_1)
=
\frac23.
\label{eq:bulk-sector-weight}
}
This identity is derived in Appendix~\ref{app:bulk-sector-weights}.
The sum of these weights exceeds one because the endpoint constraint for odd initial parity is not included.
For odd initial parity, a $t$--$0$ object must straddle the endpoint $R$, as illustrated in Fig.~\ref{fig:mapping}(b).
The next object must therefore be either $\circ$ or $\bullet_2$, which occurs with probability $1/2$.
Including this factor restores normalization: $2/3+(1/2)(2/3)=1$.

Finally, the quantity defined in Eq.~\eqref{eq:main-Gcal} satisfies
\eq{
\mathcal G_0^{(\epsilon)}(n,k)=1.
\label{eq:spin-factor-normalization}
}
Substituting Eq.~\eqref{eq:G-sector-final} into
Eq.~\eqref{eq:MGF-positive-n} and using
Eqs.~\eqref{eq:bulk-limit-p}, \eqref{eq:bulk-sector-weight}, and
\eqref{eq:spin-factor-normalization}, we obtain
\eq{
\braket{e^{\lambda Q_t}}
&=1+\frac43\sum_{n=1}^{\infty}\sum_{k=0}^{n}B_{n,k}
\Big[
p_t^{(\varnothing,\epsilon'_0)}(x_0(k),n)
\{\mathcal G_\lambda^{(0)}(n,k)-1\}
+
p_t^{(\circ,\epsilon'_1)}(x_1(k),n)
\{\mathcal G_\lambda^{(1)}(n,k)-1\}
\Big]
\\
&\quad+\frac43\sum_{n=1}^{\infty}\sum_{k=0}^{n-1}
B_{n,k}\frac{n-k}{n}
p_t^{(\bullet,\epsilon'_1)}(x_1(k),n)
\{\mathcal G_\lambda^{(1)}(n-1,k)-1\}.
\label{eq:bulk-mgf-unreduced}
}

We first consider the $r=\varnothing$ contribution in Eq.~\eqref{eq:bulk-mgf-unreduced}.
As shown in Appendix~\ref{app:bulk-generating-functions}, we have
\eq{
p_t^{(\varnothing,\epsilon)}(x,n) = \mathcal P_t^{(\epsilon)}(x,n),
\label{eq:bulk-empty-p}
}
where $\mathcal P_t^{(a)}$ is defined in Eq.~\eqref{eq:main-pt}.

We next combine the $r=\circ$ and $r=\bullet$ contributions.
In the last sum of Eq.~\eqref{eq:bulk-mgf-unreduced},
replacing $n$ by $n+1$ and using
\eq{
B_{n+1,k}\frac{n+1-k}{n+1}
=
\frac13B_{n,k}
}
makes the spin factor identical to that of the $\circ$ term.
The new $n=0$ term vanishes because $\mathcal G_\lambda^{(1)}(0,0)=1$.
The two contributions then contain the combination
\eq{
p_t^{(\circ,\epsilon)}(x,n)
+\frac13p_t^{(\bullet,\epsilon)}(x,n+1)
=
\mathcal P_t^{(1+\epsilon)}(x-1,n),
\qquad x\geq1.
\label{eq:bulk-merged-p}
}
The derivation of this identity is given in Appendix~\ref{app:bulk-generating-functions}.
We note that $x_1(k) = 0$ only when $k=0$, and $\mathcal G_\lambda^{(1)}(n,k)=1$ for $k=0$.
Hence, the $k=0$ contribution vanishes.

Combining Eqs.~\eqref{eq:bulk-empty-p} and
\eqref{eq:bulk-merged-p} gives Eq.~\eqref{eq:main-exact-mgf}.
This completes the derivation of the exact finite-time MGF in the infinite-system limit.

\subsection{Physical meaning of the exact MGF}
\label{subsec:meaning-mgf}

Let us clarify the physical meaning of the exact MGF.
Leaving aside the endpoint conditions associated with $\epsilon$ and $\epsilon'_\epsilon$, the product
$\mathcal P_t^{(\epsilon+\epsilon'_\epsilon)}(x_\epsilon(k)-\epsilon,n) B_{n,k}\mathcal G_\lambda^{(\epsilon)}(n,k)$
can be interpreted as the weighted contribution from the sector in which
the total number of $t$--$0$ charges in the observation region increases by $n$,
while the number of charges carrying the $t$--$0$ spin label $1$ increases by $k$.
In this interpretation, $\mathcal P_t^{(\epsilon+\epsilon'_\epsilon)}$
describes the probability of the charge-number change,
while $B_{n,k}$ gives the probability that exactly $k$ of these $n$ charges carry the $t$--$0$ spin label $1$.
The factor $\mathcal G_\lambda^{(\epsilon)}(n,k)$ is the conditional MGF
of the transferred magnetization in this sector.
Crucially, the time dependence enters only through $\mathcal P_t^{(\epsilon+\epsilon'_\epsilon)}$,
whereas $B_{n,k}$ and $\mathcal G_\lambda^{(\epsilon)}(n,k)$ are determined by the static spin configuration.
This structure is a direct consequence of the exact spin--charge separation.

In the original folded XXZ picture, we can interpret the above structure
of the exact MGF as follows.
As discussed below Eq.~\eqref{eq:SR-t0},
a $\circ$-object in the $t$--$0$ representation corresponds to an isolated spin,
while the ordered $t$--$0$ spin configuration describes the spin domains remaining after the isolated spins are removed.
Each $t$--$0$ spin label $2$ marks a domain wall,
and consecutive labels $1$ describe the magnetization within a domain.
Thus, $\mathcal P_t^{(\epsilon+\epsilon'_\epsilon)}$ encodes the motion
of isolated spins, which determines the portion of the static domain
configuration contributing to the transferred magnetization.
For the corresponding segment of the static domain configuration,
$B_{n,k}$ gives the statistical weight associated with $n-k$ domain walls.
The factor $\mathcal G_\lambda^{(\epsilon)}(n,k)$ then describes
the magnetization statistics obtained by summing the contributions of these domains.
Thus, in the folded XXZ picture, the dynamics is driven by isolated spins.
Domain walls do not move independently; their displacement is induced
by the motion of isolated spins and gives rise to magnetization transfer.
As shown below, this nested structure gives rise to the anomalous current fluctuations, 
and is consistent with the phenomenological argument for the folded XXZ model presented in Ref.~\cite{Gopalakrishnan_2024}.

\section{Long-time behavior of current fluctuations}
\label{sec:long-time-asymptotics}
\subsection{Long-time asymptotic analysis}
In this section, we derive the long-time scaling limit
\eq{
\lim_{t\to\infty}\braket{e^{\lambda Q_t/t^{1/4}}}
}
from the exact MGF in Eq.~\eqref{eq:main-exact-mgf}. 
The result is Eq.~\eqref{eq:main-asymptotic-mgf}.

We first observe that, for finite $n$, the definition of $\mathcal G_\lambda^{(\epsilon)}(n,k)$ gives
$
\mathcal G_{\lambda/t^{1/4}}^{(\epsilon)}(n,k)-1 =\order{t^{-1/2}}.
$
Thus, it suffices to consider large $n$ in Eq.~\eqref{eq:main-exact-mgf}.

Equation~\eqref{eq:main-exact-mgf} contains the binomial weight
\eq{
B_{n,k} = \binom nk\left(\frac23\right)^k\left(\frac13\right)^{n-k}.
}
For large $n$, Stirling's approximation gives
\eq{
B_{n,k} \simeq \frac{3}{2\sqrt{\pi n}} \exp\left[-\frac{9}{4n}\left(k-\frac{2n}{3}\right)^2\right].
\label{eq:binom-gaus}
}
Hence, the sum over $k$ is concentrated in the region
\eq{
k=\frac{2n}{3}+\order{\sqrt n}.
\label{eq:k-typical-region}
}

We next consider $\mathcal G_{\lambda}^{(\epsilon)}(n,k)$.
For $n,k\gg1$ with $k/n$ fixed in the interval $(0,1)$, the weight
$W_{n,k}(A)$ is asymptotically Gaussian in the region
$A-k/2=\order{\sqrt n}$:
\eq{
W_{n,k}(A) \simeq \sqrt{\frac{2(n-k)}{\pi kn}} \exp\left[-\frac{2(n-k)}{kn}\left(A-\frac{k}{2}\right)^2\right].
\label{eq:W-gaussian}
}
The correction $d_\epsilon(n,k)$ is of order unity and therefore does not contribute at leading order when
$\lambda$ is replaced by $\lambda/t^{1/4}$.
Replacing the sum over $A$ by a Gaussian integral gives
\eq{
\mathcal G_{\lambda/t^{1/4}}^{(\epsilon)}(n,k) \simeq \exp\left[\frac{\lambda^2}{8\sqrt t}\frac{kn}{n-k}\right].
}
Using Eq.~\eqref{eq:k-typical-region}, we obtain
\eq{
\mathcal G_{\lambda/t^{1/4}}^{(\epsilon)}(n,k) \simeq \exp\left(\frac{\lambda^2 n}{4\sqrt t}\right).
\label{eq:G-asymptotic}
}

We next need the long-time behavior of $\mathcal P_t^{(a)}(x,n)$.
For $x,n=\order{\sqrt t}$ and $a=0,1,2$, we obtain
\eq{
\mathcal P_t^{(a)}(x,n)
\simeq
\frac{2^{1-a}}{\sqrt{3t}}
\exp\left[-\frac{4\pi}{3t}\left(n-\frac{3x}{4}\right)^2\right].
\label{eq:charge-general-asymptotic}
}
Its derivation is given in Appendix~\ref{app:charge-asymptotics}.
For the arguments appearing in Eq.~\eqref{eq:main-exact-mgf},
Eqs.~\eqref{eq:def-x} and \eqref{eq:k-typical-region} give
\eq{
x_\epsilon(k)-\epsilon=\frac n3+\order{\sqrt n}.
}
Thus, on the scale $n=\order{\sqrt t}$,
\eq{
\mathcal P_t^{(\epsilon+\epsilon'_\epsilon)} (x_\epsilon(k)-\epsilon,n)
\simeq
\frac{2^{1-\epsilon-\epsilon'_\epsilon}}{\sqrt{3t}} \exp\left(-\frac{3\pi n^2}{4t}\right).
\label{eq:pt-asymptotic}
}

Substituting Eqs.~\eqref{eq:binom-gaus}, \eqref{eq:G-asymptotic}, and
\eqref{eq:pt-asymptotic} into Eq.~\eqref{eq:main-exact-mgf}, we obtain
\eqnn{
\braket{e^{\lambda Q_t/t^{1/4}}}
\simeq
1+\frac43
\sum_{\epsilon=0}^{1}\sum_{n=1}^{\infty}\sum_{k=\epsilon}^{n}
\frac{3}{2\sqrt{\pi n}}
e^{-\frac{9}{4n}(k-2n/3)^2}
\frac{2^{1-\epsilon-\epsilon'_\epsilon}}{\sqrt{3t}}
e^{-3\pi n^2/(4t)}
\left(e^{\lambda^2n/(4\sqrt t)}-1\right).
}
Note that the leading contribution comes from $n=O(\sqrt t)$ and $k=2n/3+O(\sqrt n)$.

We now perform the sums over $k$ and $\epsilon$.
The sum of the parity-dependent prefactors is
\eq{
\sum_{\epsilon=0}^{1}2^{1-\epsilon-\epsilon'_\epsilon}
=
\begin{cases}
5/2, & k \text{ even},\\
2, & k \text{ odd}.
\end{cases}
}
Since the even and odd values of $k$ carry equal weight at leading order
under the Gaussian approximation to $B_{n,k}$, we have
\eq{
\sum_{\epsilon=0}^{1}\sum_{k=0}^{n}
\frac{3}{2\sqrt{\pi n}}
e^{-\frac{9}{4n}(k-2n/3)^2}
\,2^{1-\epsilon-\epsilon'_\epsilon}
\simeq
\frac12\left(\frac52+2\right)
=
\frac94.
}
Consequently,
\eq{
\braket{e^{\lambda Q_t/t^{1/4}}}
\simeq
1+\frac{\sqrt3}{\sqrt t}\sum_{n=1}^{\infty}
e^{-3\pi n^2/(4t)}
\left(e^{\lambda^2n/(4\sqrt t)}-1\right).
\label{eq:mgf-asymptotic-sum}
}

Introducing $y=n/\sqrt t$ and replacing the sum by an integral, we find
\eq{
\lim_{t\to\infty}\braket{e^{\lambda Q_t/t^{1/4}}}
=
1+\sqrt3\int_0^\infty\dd y\,
e^{-3\pi y^2/4}\left(e^{\lambda^2y/4}-1\right).
}
Since
\eq{
\sqrt3\int_0^\infty\dd y\,e^{-3\pi y^2/4}=1,
}
this becomes
\eq{
\lim_{t\to\infty}\braket{e^{\lambda Q_t/t^{1/4}}}
=
\sqrt3\int_0^\infty\dd y\,
\exp\left(-\frac{3\pi}{4}y^2+\frac{\lambda^2}{4}y\right),
\label{eq:mgf-asym}
}
which is Eq.~\eqref{eq:main-asymptotic-mgf}.

\subsection{Physical meaning of the long-time behavior of the MGF}
The asymptotic analysis clarifies how the nested structure discussed
in Sec.~\ref{subsec:meaning-mgf} gives rise to the anomalous current fluctuations.
The motion of isolated spins determines the portion of the static domain
configuration contributing to the transferred magnetization.
Its size, measured by $n$, fluctuates on the scale $\sqrt t$,
with a Gaussian weight arising from $\mathcal P_t^{(a)}$.
For a given $n$, averaging over the domain configurations gives
Gaussian magnetization fluctuations with variance $n/2$ at leading order, as described by Eq.~\eqref{eq:G-asymptotic}.
The typical magnetization fluctuation is therefore of order $\sqrt n$, giving the current scale $t^{1/4}$.
Averaging over $n$ produces a mixture of zero-mean Gaussians
whose variance is proportional to the magnitude of another Gaussian.
This yields the non-Gaussian fluctuations represented by the limiting MGF in Eq.~\eqref{eq:mgf-asym}.

\section{Application to a spin-spin correlation function}
\label{sec:spin-correlation}

Our approach also applies to the spin-spin dynamical correlation function.
To this end, we introduce the spatially extended MGF
\eq{
M_t(\lambda; \ell) :=
\Tr\left[
e^{\lambda S_{\leq \ell}}e^{-iHt} e^{-\lambda S_{\leq 0}}\rho e^{iHt}
\right],
\qquad
S_{\leq \ell}:=\sum_{j\leq \ell }s_j^z.
\label{eq:spatial-mgf}
}
In particular, we have $M_t(\lambda;0)=\braket{e^{\lambda Q_t}}$ and 
\eq{
\braket{s^z_\ell (t) s^z_0} = \frac12\partial^2_{\lambda}\big[ M_t(\lambda; \ell+1) +M_t(\lambda; \ell-1) -2 M_t(\lambda; \ell) \big]|_{\lambda=0}.
}

For an even displacement $\ell=2m$, with $m\in\mathbb Z$, the length constraint in Eq.~\eqref{eq:length-constraint}
and the sector decomposition in Sec.~\ref{subsec:decomp} retain essentially the same structure as for $\ell=0$.
For an odd displacement $\ell=2m-1$, these require minor modifications.
For simplicity, we restrict our analysis to even displacements.
The dynamical correlation function of spins summed over two adjacent sites can be obtained from $M_t(\lambda ; 2m)$ through
\eq{
&\langle (s^z_{2m}(t)+s^z_{2m-1}(t)) (s^z_0+s^z_{-1}) \rangle 
\\
&\quad =
\frac12\partial_\lambda^2
\big[ M_t(\lambda;2m+2)-2M_t(\lambda;2m)+M_t(\lambda;2m-2) \big] |_{\lambda=0}.
\label{eq:block-correlation}
}
Hereafter, we derive the long-time asymptotic form of this correlation function.

Following the same steps as for $\ell=0$, we obtain
\eq{
M_t(\lambda;2m)
&=1+\frac23 \sum_{\epsilon=0}^{1}\sum_{n=1}^{\infty}\sum_{k=\epsilon}^{n} B_{n,k}\left\{\mathcal G_\lambda^{(\epsilon)}(n,k)-1\right\}
\\
&\times\left[
\mathcal P_t^{(\epsilon+\epsilon'_\epsilon)}
\left(m+x_\epsilon(k)-\epsilon,n\right)
+
\mathcal P_t^{(\epsilon+\epsilon'_\epsilon)}
\left(-m+x_\epsilon(k)-\epsilon,n\right)
\right].
\label{eq:spatial-mgf-even}
}
Here, the quantities on the right-hand side are those defined in Sec.~\ref{sec:main-results}.
While $\mathcal P_t^{(a)}(x,n)$ in Eq.~\eqref{eq:main-pt} is defined for $x\in\mathbb Z_{\geq0}$,
Eq.~\eqref{eq:spatial-mgf-even} also involves negative spatial arguments.
To extend its definition, we first extend $F_t$ by
\eq{
F_t(\theta; x) := F_t(-\theta; -x), 
\qquad x\in\mathbb Z_{<0}.
}
We then define $\mathcal P_t^{(a)}(x,n)$ for $x<0$ by applying Eqs.~\eqref{eq:main-pt} and \eqref{eq:main-difference-operator} 
to this extended function.

A similar asymptotic analysis gives the long-time asymptotic form of the rescaled MGF,
\eq{
M_t(\lambda/t^{1/4};2m)
\simeq
\frac{\sqrt3}{2}\int_{-\infty}^{\infty}\dd y\,
\exp\left[
-\frac{3\pi y^2}{4}
+\frac{\lambda^2}{4}\left|\frac{m}{\sqrt t}+y \right|
\right],
}
where $m/\sqrt t$ is held fixed as $t\to\infty$.
Substituting this expression into Eq.~\eqref{eq:block-correlation}
and replacing the spatial second difference by the corresponding derivative, we obtain
\eq{
\left\langle
\bigl(s^z_{2m}(t)+s^z_{2m-1}(t)\bigr)
\bigl(s^z_0+s^z_{-1}\bigr)
\right\rangle
\simeq
\frac{\sqrt3}{4\sqrt t}
\exp\left[-\frac{3\pi m^2}{4t}\right].
}
Thus, the dynamical correlation function exhibits a diffusive Gaussian profile at long times,
with diffusion constant $D=4/(3\pi)$~\cite{Gopalakrishnan_2019,Feldmeier_2022}
when distance is measured in units of the original lattice spacing.
This Gaussian profile agrees with the results
previously obtained via the analysis of tracer particles~\cite{Feldmeier_2022}.

\section{Conclusion}
\label{sec:conclusion}
In this work, we have derived an exact expression for the moment-generating function of the time-integrated spin current
in the folded XXZ spin chain at infinite temperature.
The result is expressed in terms of a static combinatorial weight and a Fredholm determinant associated with the continuous Bessel kernel.
From its long-time asymptotics, we have shown that the typical fluctuations grow on the scale $t^{1/4}$
and that the rescaled current converges to the non-Gaussian distribution described by the M-Wright function.
This asymptotic behavior agrees with the strong-anisotropy limit of 
the hydrodynamic prediction for the XXZ chain~\cite{Yoshimura_2026}.
Moreover, by extending our analysis, we have obtained a diffusive Gaussian profile
for the dynamical correlation function of spins summed over two adjacent sites.

A key ingredient of the derivation is the mapping to the one-dimensional $t$--$0$ model~\cite{Dias_2000,Pozsgay_2021_1,Feldmeier_2022},
whose dynamics exhibits exact spin--charge separation~\cite{Ogata_1990,Izergin_1998,Gamayun_2023,Gamayun_2024}.
Exploiting this mapping, we reduce the folded XXZ dynamics to two contributions:
a dynamical contribution from isolated spins and a static contribution from spin domains.
Their interplay produces the non-Gaussian distribution described by the M-Wright function, 
providing a microscopic explanation of the anomalous current fluctuations in the folded XXZ chain.

Several directions remain open.
A natural next step is to extend the exact microscopic analysis to the XXZ chain at finite anisotropy,
where the spin--charge separation underlying the present solution is no longer directly available.
It would also be interesting to apply the present approach to spatially inhomogeneous initial states.

\section*{Acknowledgements}
We thank \v{Z}iga Krajnik and Toma\v{z} Prosen for helpful comments.
TS would also like to thank the Center of Mathematical Sciences and Applications at Harvard University for the hospitality and stimulating environment during the program `Classical, Quantum, and Probabilistic Integrable Systems', where he had initial discussions on the problem with them. 
\paragraph{Funding information}
The work of TI has been supported by JST SPRING, Japan Grant Number JPMJSP2180.
The work of KF has been supported by JSPS KAKENHI Grant No. JP23K13029.
The work of TS was supported by JSPS KAKENHI Grant Numbers JP21H04432 and JP23K22414.

\appendix

\section{Numerical calculation of the characteristic function}
\label{app:numerics}

We evaluate the rescaled characteristic function $\braket{e^{i \theta Q_t /t^{1/4}}}$
using matrix product operator (MPO) simulations of an integrable Trotterized circuit of the folded XXZ model~\cite{Pozsgay_2021_2,Gombor_2021}.
We consider an even number $L$ of sites, $j=-L/2+1,\ldots,L/2$, with open boundary conditions,
and place the measurement cut between sites $0$ and $1$.
In this appendix, $L$ denotes the physical chain length, rather than the number of $t$--$0$ objects introduced in
Sec.~\ref{sec:mapping}.
The magnetization to the left of the cut is
\eq{
S_{\rm left}:=\sum_{j=-L/2+1}^{0}s_j^z.
}

\subsection{Integrable circuit}

Let $U_j(u)$ denote the local gate acting on sites $j,\ldots,j+3$,
where $u$ is the circuit parameter.
When the two outer spins are equal, the gate acts on the central
antiparallel spins as
\eqs{
U_j(u)\ket{\sigma,\uparrow,\downarrow,\sigma}
&=
\frac{1}{\cosh u}\ket{\sigma,\uparrow,\downarrow,\sigma}
+i\tanh u\ket{\sigma,\downarrow,\uparrow,\sigma},
\\
U_j(u)\ket{\sigma,\downarrow,\uparrow,\sigma}
&=
\frac{1}{\cosh u}\ket{\sigma,\downarrow,\uparrow,\sigma}
+i\tanh u\ket{\sigma,\uparrow,\downarrow,\sigma},
}
where $\sigma\in\{\uparrow,\downarrow\}$.
The gate acts as the identity on all other spin configurations.

For $r=0,1,2$, define the layers
\eq{
V_r(u):=\prod_{\substack{j\equiv r \bmod3\\
-L/2+1\leq j\leq L/2-3}}U_j(u).
}
The gates within each layer commute.
The implementation applies the layers in the order $r=2,0,1$,
so that one circuit period is
\eq{
V(u):=V_1(u)V_0(u)V_2(u).
}
For $u \ll 1$, the local gate satisfies 
\eq{
U_j(u)
=
I+iu(s_{j+1}^+s_{j+2}^-+s_{j+2}^+s_{j+1}^-)
\left(2s_j^zs_{j+3}^z+\frac12\right)
+\order{u^2}.
}
Hence, this circuit can be regarded as the Trotterization of the folded XXZ model:
\eq{
V(u)=I-2iuH+\order{u^2}.
}
We therefore assign the physical time
\eq{
t=2um
}
to $m$ circuit periods.
The continuous-time Hamiltonian dynamics is recovered by taking $u\to0$ and $m \to \infty$ at fixed $t = 2um$.

\subsection{MPO evaluation}

Define the tilted circuit
\eq{
V_\theta(u):=e^{i\theta S_{\rm left}}V(u)e^{-i\theta S_{\rm left}}.
}
At infinite temperature, the circuit characteristic function is
\eq{
\braket{e^{i\theta Q_t}}_{L,u}
=
2^{-L}\Tr\left[V_\theta(u)^m\bigl(V(u)^\dagger\bigr)^m\right],
\qquad t=2um.
}
We evaluate this trace by representing
\eq{
W_m:=V_\theta(u)^m\bigl(V(u)^\dagger\bigr)^m
}
as an MPO and updating it according to
\eq{
W_{m+1}=V_\theta(u)W_mV(u)^\dagger,
\qquad W_0=I.
}

Each local update takes the form
\eq{
W\longmapsto U_{j,\theta}(u)W U_j(u)^\dagger,
}
where $U_{j,\theta}(u)=U_j(u)$ unless $j=-1$.
Only the gate starting at $j=-1$ exchanges spins across the measurement cut.
For this gate,
\eqs{
U_{-1,\theta}(u)
=
e^{i\theta(s_{-1}^z+s_0^z)}
U_{-1}(u)
e^{-i\theta(s_{-1}^z+s_0^z)}.
}
Thus, the counting field modifies only the gate that transports magnetization across the cut.

After each four-site update, we restore the MPO form by three successive singular-value decompositions.
We impose a maximum bond dimension $\chi$ and an SVD truncation cutoff of $10^{-10}$
for the relative sum of the squared discarded singular values.
No additional normalization is performed after truncation.
For numerical convenience, we represent $W_m/\sqrt{2^L}$ as an MPO.
Contracting it with $I/\sqrt{2^L}$ gives the normalized trace $2^{-L}\Tr W_m$.

\subsection{Parameters and convergence checks}

The results in Fig.~\ref{fig:current-scaling} use $L=128$, $u=0.1$, and $\chi=128$.
We evaluate the characteristic function at physical times $t=4,8,16$, corresponding to $m=20,40,80$ circuit periods.
For each time, the unscaled counting field is sampled at $0.1,0.2,\ldots,2.0$.
The horizontal coordinate in Fig.~\ref{fig:current-scaling}
is obtained by multiplying these values by $t^{1/4}$,
so that the plotted quantity is $\operatorname{Re}\braket{e^{i\theta Q_t/t^{1/4}}}$.

\begin{figure}[t]
\centering
\includegraphics[width=0.9\linewidth]{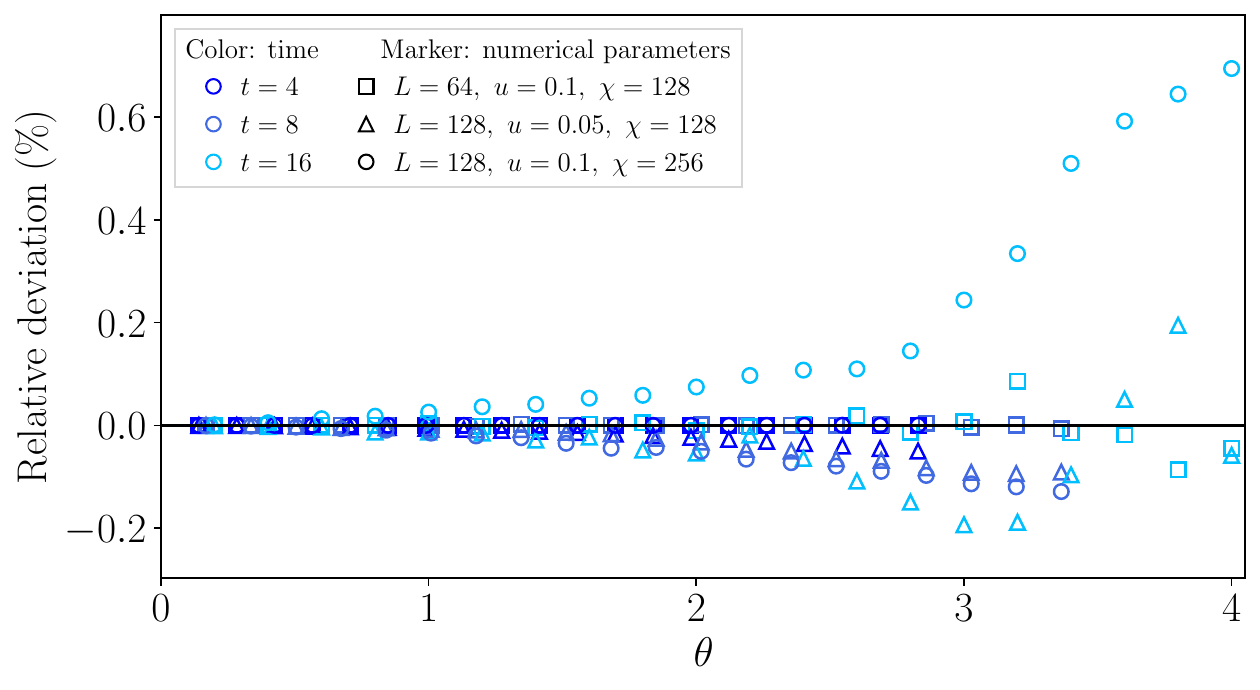}
\caption{
Relative deviations of the real part of the characteristic function
$\braket{e^{i\theta Q_t /t^{1/4}}}$ from the reference calculation with $(L,u,\chi)=(128,0.1,128)$.
The deviations are defined as the difference between each numerical
result and the reference, divided by the reference and expressed as a percentage.
Colors indicate the physical times $t=4,8,16$,
while markers distinguish the three parameter sets specified in the legend.
The absolute relative deviations remain below approximately $0.7\%$ throughout the plotted range.
}
\label{fig:numerical-convergence}
\end{figure}

To assess numerical convergence, we compare the four parameter sets
\eqs{
(L,u,\chi) =(64,0.1,128),\quad(128,0.1,128),\quad (128,0.05,128),\quad (128,0.1,256).
}
We use $(L,u,\chi)=(128,0.1,128)$, employed in Fig.~\ref{fig:current-scaling}, as the reference.
Figure~\ref{fig:numerical-convergence} shows the relative deviations
of the real part of the characteristic function from this reference at $t=4,8,16$.

The comparison between $L=64$ and $L=128$ at fixed $u=0.1$ and $\chi=128$ tests the system-size dependence.
Similarly, comparing $u=0.1$ and $u=0.05$ at fixed $L=128$ and $\chi=128$ tests the dependence on the circuit parameter,
while comparing $\chi=128$ and $\chi=256$ at fixed $L=128$ and $u=0.1$ tests the dependence on the MPO truncation.
The absolute relative deviations from the reference remain below
approximately $0.7\%$ over the time and counting-field ranges shown,
supporting the numerical convergence of the results in Fig.~\ref{fig:current-scaling}.

\section{Derivation of the sector contribution}
\label{app:derivation-G-sector}

We derive the expression for $G(\lambda\mid\omega';\omega)$ in Eq.~\eqref{eq:G-sector-final}.
As in the main text, for $N'-N>0$, we use the notation
\eq{
n =N'-N>0, \qquad k =N_1'-N_1, \qquad 0\leq k\leq n,
}
and
\eq{
\epsilon_{\rm i} = \frac{1-(-1)^{N_1}}{2}, \qquad \epsilon_{\rm f} = \frac{1-(-1)^{N_1'}}{2} = (\epsilon_{\rm i}+k)\bmod2.
}

\subsection{Spin traces}
\label{app:sector-spin-traces}

Expanding the endpoint projectors in Eq.~\eqref{eq:G-spin-charge} and
tracing out the spin degrees of freedom on sites
$N'+1,N'+2,\ldots$, we obtain
\eqs{
G(\lambda\mid\omega';\omega)
={}&\frac12\sum_{\sigma=\pm1}\mathcal P_c(L,N)
\begin{cases}
p_t^{(\varnothing,\epsilon_{\rm f})}(L',N';L,N)\mathcal S_{\omega',\omega}^{(\varnothing)}(\sigma\lambda), & \epsilon_{\rm i}=0,\\
p_t^{(\circ,\epsilon_{\rm f})}(L',N';L,N)\mathcal S_{\omega',\omega}^{(\circ)}(\sigma\lambda), & \epsilon_{\rm i}=1,
\end{cases}
\notag\\
&+\frac12\sum_{\sigma=\pm1}\mathcal P_c(L,N)
\begin{cases}
0, & \epsilon_{\rm i}=0,\\
p_t^{(\bullet,\epsilon_{\rm f})}(L',N';L,N)\mathcal S_{\omega',\omega}^{(\bullet)}(\sigma\lambda), & \epsilon_{\rm i}=1.
\end{cases}
\label{eq:app-G-general}
}
For $r\in\{\varnothing,\circ,\bullet\}$, define
\eq{
S_j^{(\varnothing)} := I_s,\qquad
S_j^{(\circ)} := I_s,\qquad
S_j^{(\bullet)} := \ket{2}\bra{2}_j.
}
We then define
\eq{
\mathcal S_{\omega',\omega}^{(r)}(\mu)
:=
\Tr_s \bigg[
P_{N_1'}^{N'}(\mu)P_{N_1}^{N}(-\mu)S_{N+1}^{(r)}
\bigotimes_{j=1}^{N'}
\Big[\frac23\ket{1}\bra{1}+\frac13\ket{2}\bra{2}\Big]_j
\bigg].
\label{eq:app-spin-factor-definition}
}
Note that the definition of $p_t^{(r,\epsilon)}(L',N';L,N)$ is given in Eq.~\eqref{eq:charge-factor-main}.

We evaluate the quantities
\eq{
\frac12\sum_{\sigma=\pm1} \mathcal S_{\omega',\omega}^{(r)}(\sigma\lambda).
\label{eq:app-spin-factor-average}
}

We first consider $r\in\{\varnothing,\circ\}$.
The two cases give the same result.
We denote the additional $t$--$0$ spin variables contained in the final
sector by
\eq{
\xi_j:=\eta_{s,N+j}, \qquad j=1,\ldots,n.
}
Since the initial and final sectors contain $N_1$ and
$N_1'=N_1+k$ spin labels $1$, respectively, the additional sequence
$\bm\xi=(\xi_1,\ldots,\xi_n)$ contains exactly $k$ labels $1$:
\eq{
\sum_{j=1}^{n}\ind{\xi_j=1}=k.
\label{eq:app-xi-constraint}
}

For later convenience, we define
\eq{
A_n(\bm\xi) := \sum_{j=1}^{n} (-1)^{\sum_{\ell<j}\ind{\xi_\ell=2}} \ind{\xi_j=1}.
\label{eq:app-An}
}
Using the definition of $F_{N,N_1}$ in
Eq.~\eqref{eq:def-F-sector}, we obtain
\eq{
F_{N',N_1'}(\eta_{s,1},\ldots,\eta_{s,N},\bm\xi)
-F_{N,N_1}(\eta_{s,1},\ldots,\eta_{s,N})
=
(-1)^{N-N_1}
\left[\frac12A_n(\bm\xi)+d_{\epsilon_{\rm i}}(n,k)\right].
\label{eq:app-F-difference}
}
Here,
\eq{
d_{\epsilon}(n,k) := \frac14(-1)^{n-k} \left[1+(-1)^{\epsilon+k}\right] - \frac14\left[1+(-1)^{\epsilon}\right].
\label{eq:app-d-epsilon}
}

The first $N$ spin variables contribute only through their sector
weight
\eq{
\Tr_s\left[P_{N_1}^{N}\rho_s\right] = \mathcal P_s(N,N_1) = \binom{N}{N_1} \left(\frac23\right)^{N_1} \left(\frac13\right)^{N-N_1}.
\label{eq:app-Ps}
}
For a fixed sequence $\bm\xi$ containing $k$ labels $1$, its
probability under $\rho_s$ is
\eq{
\left(\frac23\right)^k \left(\frac13\right)^{n-k}.
}
Therefore, for $r\in\{\varnothing,\circ\}$,
\eq{
\frac12\sum_{\sigma=\pm1}
\mathcal S_{\omega',\omega}^{(r)}(\sigma\lambda)
=
\mathcal P_s(N,N_1)
\left(\frac23\right)^k
\left(\frac13\right)^{n-k}
\sum_{\substack{
\bm\xi\in\{1,2\}^n\\
\sum_{j=1}^{n}\ind{\xi_j=1}=k
}}
\ch\left[
\lambda\left\{
\frac12A_n(\bm\xi)
+
d_{\epsilon_{\rm i}}(n,k)
\right\}
\right].
\label{eq:app-spin-factor-unconstrained}
}
Here, the overall factor $(-1)^{N-N_1}$ in
Eq.~\eqref{eq:app-F-difference} drops out because the average over
$\sigma=\pm1$ produces the hyperbolic cosine.

Introducing
\eq{
B_{n,k} := \binom nk \left(\frac23\right)^k \left(\frac13\right)^{n-k},
\label{eq:app-Bnk}
}
and the normalized combinatorial factor
\eq{
\mathcal G_{\lambda}^{(\epsilon)}(n,k)
:=
\frac{1}{\binom nk}
\sum_{\substack{
\bm\xi\in\{1,2\}^n\\
\sum_{j=1}^{n}\ind{\xi_j=1}=k
}}
\ch\left[
\lambda\left\{
\frac12A_n(\bm\xi)
+
d_{\epsilon}(n,k)
\right\}
\right],
\label{eq:app-Gcal}
}
Eq.~\eqref{eq:app-spin-factor-unconstrained} becomes
\eq{
\frac12\sum_{\sigma=\pm1} \mathcal S_{\omega',\omega}^{(\varnothing)}(\sigma\lambda)
=
\frac12\sum_{\sigma=\pm1} \mathcal S_{\omega',\omega}^{(\circ)}(\sigma\lambda)
=
\mathcal P_s(N,N_1)B_{n,k} \mathcal G_{\lambda}^{(\epsilon_{\rm i})}(n,k).
\label{eq:app-spin-factor-unconstrained-final}
}

We next consider the case $r=\bullet$.
In this case, the initial endpoint projector contains
\eq{
S_{N+1}^{(\bullet)} = \ket{2}\bra{2}_{N+1},
}
and therefore imposes the additional constraint
\eq{
\xi_1=\eta_{s,N+1}=2.
\label{eq:app-xi1-two}
}
The remaining sequence $(\xi_2,\ldots,\xi_n)$ has length $n-1$ and
contains exactly $k$ labels $1$ and $n-k-1$ labels $2$.

For a fixed constrained sequence $\bm\xi$, the probability under
$\rho_s$ is again
\eq{
\left(\frac23\right)^k \left(\frac13\right)^{n-k}.
}
Hence,
\eqnn{
\frac12\sum_{\sigma=\pm1}
\mathcal S_{\omega',\omega}^{(\bullet)}(\sigma\lambda)
&=
\mathcal P_s(N,N_1)
\left(\frac23\right)^k \left(\frac13\right)^{n-k}
\sum_{\substack{
\bm\xi\in\{1,2\}^n\\ \xi_1=2\\
\sum_{j=1}^{n}\ind {\xi_j=1}=k
}}
\ch\left[
\lambda\left\{
\frac12A_n(\bm\xi) +d_{1}(n,k)
\right\}
\right].
\label{eq:app-spin-factor-constrained}
\\
&=
\mathcal P_s(N,N_1)
\left(\frac23\right)^k \left(\frac13\right)^{n-k}
\sum_{\substack{
\bm\xi\in\{1,2\}^{n-1}\\ 
\sum_{j=1}^{n-1}\ind {\xi_j=1}=k
}}
\ch\left[
\lambda\left\{
\frac12A_{n-1}(\bm\xi) +d_{1}(n-1,k)
\right\}
\right].
}
Here we used $\epsilon_{\rm i} =1$ for $r=\bullet$ and 
\eq{
A_n(2,\xi_2,\cdots,\xi_n) = - A_{n-1}(\xi_2,\cdots,\xi_n),\qquad d_1(n,k) = -d_1(n-1,k).
}
Thus, we obtain
\eq{
\frac12\sum_{\sigma=\pm1} \mathcal S_{\omega',\omega}^{(\bullet)}(\sigma\lambda)
=
\mathcal P_s(N,N_1)B_{n,k}\frac{n-k}{n}
\mathcal G_{\lambda}^{(1)}(n - 1,k).
\label{eq:app-spin-factor-constrained-final}
}

\subsection{Combinatorial evaluation of the spin factor}
\label{app:sector-combinatorics}

We now evaluate the combinatorial factors
$\mathcal G_{\lambda}^{(\epsilon)}(n,k)$ explicitly.
Every sequence $\bm\xi$ containing $k$ labels $1$ and $n-k$ labels
$2$ can be uniquely written as
\eq{
1^{a_1}2\,1^{a_2}2\cdots2\,1^{a_{n-k+1}},
\qquad
a_j\in\mathbb Z_{\geq0},
\qquad
\sum_{j=1}^{n-k+1}a_j=k.
}
For such a sequence, Eq.~\eqref{eq:app-An} gives
\eq{
A_n(\bm\xi) = a_1-a_2+a_3-a_4+\cdots.
}
Defining
\eq{
A := a_1+a_3+a_5+\cdots,
}
we obtain
\eq{
\frac12A_n(\bm\xi) = A-\frac{k}{2}.
\label{eq:app-A-relation}
}

For fixed $A$, the sum of the odd-indexed variables $a_j$ is $A$,
whereas that of the even-indexed variables is $k-A$.
The number of sequences corresponding to a given $A$ is therefore
\eq{
\binom{A+\lfloor(n-k)/2\rfloor}{\lfloor(n-k)/2\rfloor}
\binom{k-A+\lceil(n-k)/2\rceil-1}{\lceil(n-k)/2\rceil-1}.
}
We consequently define the normalized weight
\eq{
W_{n,k}(A)
:=
\frac{1}{\binom nk}
\binom{A+\lfloor(n-k)/2\rfloor}{\lfloor(n-k)/2\rfloor}
\binom{k-A+\lceil(n-k)/2\rceil-1}{\lceil(n-k)/2\rceil-1}.
\label{eq:app-Wnk}
}
We use the convention
\eq{
\binom{-1}{-1}=1, \qquad \binom{r}{-1}=0 \quad(r\neq-1),
}
so that Eq.~\eqref{eq:app-Wnk} also applies to $n=k$.
Substituting Eqs.~\eqref{eq:app-A-relation} and~\eqref{eq:app-Wnk} into
Eq.~\eqref{eq:app-Gcal}, we obtain
\eq{
\mathcal G_{\lambda}^{(\epsilon)}(n,k)
=
\sum_{A=0}^{k}W_{n,k}(A)
\ch\left[
\lambda\left\{A-\frac{k}{2}+d_{\epsilon}(n,k)\right\}
\right],
}
which is Eq.~\eqref{eq:main-Gcal}.

Finally, substituting
Eqs.~\eqref{eq:app-spin-factor-unconstrained-final} and
\eqref{eq:app-spin-factor-constrained-final} into
Eq.~\eqref{eq:app-G-general}, we obtain, for $\epsilon_{\rm i}=0$,
\eq{
G(\lambda\mid\omega';\omega)
=
\mathcal P_c(L,N)\mathcal P_s(N,N_1)B_{n,k}
p_t^{(\varnothing,\epsilon_{\rm f})}(L',N';L,N)
\mathcal G_{\lambda}^{(0)}(n,k),
}
whereas for $\epsilon_{\rm i}=1$,
\eqs{
G(\lambda\mid\omega';\omega)
={}&
\mathcal P_c(L,N)\mathcal P_s(N,N_1)B_{n,k}
\Big[
p_t^{(\circ,\epsilon_{\rm f})}(L',N';L,N)
\mathcal G_{\lambda}^{(1)}(n,k)
\notag\\
&\hspace{29mm}
+\frac{n-k}{n}p_t^{(\bullet,\epsilon_{\rm f})}(L',N';L,N)
\mathcal G_{\lambda}^{(1)}(n-1,k)
\Big].
}
This proves Eq.~\eqref{eq:G-sector-final}.

\section{Derivation of the MGF in the infinite-system limit}
\label{app:bulk-lim}

We derive the infinite-system limit of the transition weights, evaluate the
initial-sector weights, and reduce the endpoint contributions to a
single generating function, which we then express as a Fredholm determinant.
\subsection{Infinite-system limit of the transition weights}
\label{app:bulk-transition}

Here, we derive the infinite-system limit of
$p_t^{(r,\epsilon)}(L+x,N+n;L,N)$ appearing in Eq.~\eqref{eq:bulk-limit-p}.
In this subsection, $t\in \mathbb{R}_{\geq0}$, $x\in\mathbb Z_{\geq0}$, and $n\in\mathbb Z$
are held fixed while the infinite-system limit is taken.

We use the Fourier representation
\eq{
P_N^L =\int_{-\pi}^{\pi}\frac{d\theta}{2\pi} e^{-iN\theta}e^{i\theta\mathcal N_L},
\qquad
\mathcal N_L:=\sum_{j=1}^L n_j,
\label{eq:fourier_proj}
}
for the final projector $P_{N+n}^{L+x}$ in
Eq.~\eqref{eq:charge-factor-main}.
Using also
$e^{-i\theta\mathcal N_L}P_N^L=e^{-iN\theta}P_N^L$, we obtain
\eqs{
p_t^{(r,\epsilon)}(L+x,N+n;L,N)
={}&
\int_{-\pi}^{\pi}\frac{d\theta}{2\pi}e^{-in\theta}
\Tr_c\Big[
E_{L+x+1}^{\epsilon}e^{i\theta\mathcal N_{L+x}}
e^{-iH_{\xx}t}e^{-i\theta\mathcal N_L}C_{L+1}^{(r)}
\notag\\
&\hspace{19mm}\times
\frac{P_N^L}{\binom LN}
\bigotimes_{j>L}
\left[\frac14\ket{\circ}\bra{\circ}+\frac34\ket{\bullet}\bra{\bullet}\right]_j
e^{iH_{\xx}t}
\Big].
}

Let $O$ be an operator supported on a fixed finite set of sites.
Then, after shifting the coordinate by $L$, 
\eq{
\lim_{\substack{L,N\to\infty\\N/L\to3/4}}
\Tr_c\left[
O\,\frac{P_N^L}{\binom LN}
\bigotimes_{j>L}\left(\frac14\ket{\circ}\bra{\circ}+\frac34\ket{\bullet}\bra{\bullet}\right)_j
\right]
=
\Tr_c\left[
O\,
\bigotimes_{j\in\mathbb Z}
\left(\frac14\ket{\circ}\bra{\circ}+\frac34\ket{\bullet}\bra{\bullet}\right)_j
\right],
\label{eq:canonical-bernoulli-limit}
}
Indeed, on any fixed set of sites, the fixed-charge distribution converges to the Bernoulli distribution with density $3/4$.

The operator to which this limit is applied is
\eq{
e^{iH_{\xx}t} E_{L+x+1}^{\epsilon} e^{i\theta\mathcal N_{L+x}} e^{-iH_{\xx}t} e^{-i\theta\mathcal N_L} C_{L+1}^{(r)}.
}
At $t=0$, these factors have finite support because
\eq{
e^{i\theta\mathcal N_{L+x}}e^{-i\theta\mathcal N_L} = e^{i\theta\sum_{j=L+1}^{L+x}n_j}.
}
For fixed $t$, locality of the Hamiltonian implies that the time-evolved
operator can be approximated arbitrarily well by an operator with finite
support.
Equation~\eqref{eq:canonical-bernoulli-limit} can therefore be applied.
After the coordinate shift, we use $H_{\xx}$ for the XX Hamiltonian on
the full lattice,
\eq{
H_{\xx}
:=
-\frac12\sum_{j\in\mathbb Z}
\left(\ket{\circ\bullet}\bra{\bullet\circ}+\ket{\bullet\circ}\bra{\circ\bullet}\right)_{j,j+1},
\label{eq:bulk-XX-Hamiltonian}
}
and $\rho_c$ for the translation-invariant product state
\eq{
\rho_c
:=
\bigotimes_{j\in\mathbb Z}
\left[\frac14\ket{\circ}\bra{\circ}+\frac34\ket{\bullet}\bra{\bullet}\right]_j.
\label{eq:bulk-charge-state}
}
For each integer $x$, we also define the left-half charge
\eq{
\mathcal N_x:=\sum_{j\leq x}n_j.
\label{eq:bulk-left-charge}
}
With this notation, we obtain
\eq{
\lim_{\substack{L,N\to\infty\\N/L\to3/4}}
p_t^{(r,\epsilon)}(L+x,N+n;L,N)
=
\int_{-\pi}^{\pi}\frac{d\theta}{2\pi}e^{-in\theta}
F_t^{(r,\epsilon)}(\theta;x),
\label{eq:app-bulk-F}
}
where $F_t^{(r,\epsilon)}(\theta;x)$ is defined in Eq.~\eqref{eq:def-F}.
This proves Eq.~\eqref{eq:bulk-limit-p}.

\subsection{Initial-sector weights}
\label{app:bulk-sector-weights}
\label{app:sum-ini}

We evaluate the sum of the initial-sector weights $\mathcal P_c(L,N)\mathcal P_s(N,N_1)$.
Since the initial-sector weights are given in Eq.~\eqref{eq:initial-weights}, we have
\eqnn{
\sum_{\substack{\omega \in \Omega  \\ N_1 \in 2\mathbb{Z}_{\geq 0} +\epsilon}  } \mathcal P_c(L,N)\mathcal P_s(N,N_1)  
= \sum_{\substack{\omega \in \Omega  \\ N_1 \in 2\mathbb{Z}_{\geq 0} + \epsilon} } \frac{L!}{(L-N_1)! N_1! } \frac{(L-N_1)!}{(L-N)! (N-N_1)!} 
\left(\frac 1 2\right)^{L-N_1} \left(\frac 1 2\right)^{L}.
}
By summing over $N$ with fixed $L,N_1$, we obtain
\eq{
\sum_{\substack{\omega \in \Omega  \\ N_1 \in 2\mathbb{Z}_{\geq 0} + \epsilon } } \mathcal P_c(L,N)\mathcal P_s(N,N_1)  = 
\sum_{\substack{L \geq N_1 \geq 0 \\ 2L - N_1 = R - \epsilon }}  \frac{L!}{(L-N_1)! N_1! } \left(\frac 1 2\right)^{L}.
}

To evaluate the above quantity, we define
\eq{
\mathcal{A}_l :=  \sum_{\substack{L \geq N_1 \geq 0 \\ 2L - N_1 = l }}  \frac{L!}{(L-N_1)! N_1! } \left(\frac 1 2\right)^{L}
}
for $l \geq 0$. By setting $D = L-N_1$,  we can calculate its generating function
\eqsnn{
\sum_{l \geq 0} \mathcal A_l z^l 
&= \sum_{l \geq 0}z^l \sum_{\substack{D, N_1 \geq 0 \\ 2D + N_1 = l }}  \frac{(D+N_1)!}{D! N_1! } \left(\frac 1 2\right)^{D+N_1}
\\
&= \sum_{\substack{D, N_1 \geq 0 }}  \frac{(D+N_1)!}{D! N_1! } \left(\frac{ z^2}{ 2}\right)^{D} \left(\frac {z} {2} \right)^{N_1}
\\
&= \frac{1}{1 - z/2 - z^2/2}.
}

Using
\eq{
\frac{1}{1-z/2-z^2/2} = \frac{2/3}{1-z} + \frac{1/3}{1+z/2},
}
we obtain
\eq{
\mathcal A_\ell=\frac23+\frac13\left(-\frac12\right)^\ell.
\label{eq:app-AL}
}
Therefore, it follows that
\eq{
\sum_{\substack{\omega\in\Omega\\N_1\in2\mathbb Z_{\geq 0}}}\mathcal P_c(L,N)\mathcal P_s(N,N_1)
=
\mathcal A_R,
\qquad
\sum_{\substack{\omega\in\Omega\\N_1\in2\mathbb Z_{\geq 0} + 1}}\mathcal P_c(L,N)\mathcal P_s(N,N_1)
=
\mathcal A_{R-1}.
}
Taking $R\to\infty$ in Eq.~\eqref{eq:app-AL} gives
\eq{
\lim_{R\to\infty}
\sum_{\substack{\omega\in\Omega\\N_1\in2\mathbb Z_{\geq 0} + \epsilon}}
\mathcal P_c(L,N)\mathcal P_s(N,N_1)
=
\frac23,
\qquad
\epsilon=0,1,
}
which proves Eq.~\eqref{eq:bulk-sector-weight}.

\subsection{Reduction and evaluation of the generating functions}
\label{app:bulk-generating-functions}

We first show that $F_t^{(r,\epsilon)}(\theta;x)$ given in Eq.~\eqref{eq:def-F} reduces to 
$F_t^{(\varnothing,0)}(\theta;x)$.

We use
\eq{
E_j=I_c-\frac23n_j,
\qquad
e^{i\theta n_j}=I_c+(e^{i\theta}-1)n_j.
}
Recall that $E_j$ and $n_j$ are given in Eqs.~\eqref{eq:def-E-pro} and \eqref{eq:def-n}.
Then, since $\mathcal N_{x+1}=\mathcal N_x+n_{x+1}$,
\eq{
E_{x+1}e^{i\theta\mathcal N_x} =
\frac{(3e^{i\theta}-1)e^{i\theta\mathcal N_x}
-2e^{i\theta\mathcal N_{x+1}}}
{3(e^{i\theta}-1)}.
\label{eq:final-endpoint-identity}
}
Substitution into Eq.~\eqref{eq:def-F} yields
\eq{
F_t^{(r,1)}(\theta;x) = \bigl(\mathsf T_\theta F_t^{(r,0)}(\theta;\cdot)\bigr)(x),
\label{eq:final-endpoint-reduction}
}
where $\mathsf T_{\theta}$ is defined in Eq.~\eqref{eq:main-difference-operator}.
In particular, defining
\eq{
F_t(\theta;x):=F_t^{(\varnothing,0)}(\theta;x)
=
\Tr_c\left[
e^{i\theta\mathcal N_x}e^{-iH_{\xx}t}
e^{-i\theta\mathcal N_0}\rho_c e^{iH_{\xx}t}
\right],
\label{eq:basic-charge-trace}
}
we have $F_t^{(\varnothing,\epsilon)}=\mathsf T_\theta^\epsilon F_t$.

For the initial endpoint, the relevant local identity is
\eq{
e^{-i\theta\mathcal N_0}
\left(C_1^{(\circ)}+\frac{e^{-i\theta}}3C_1^{(\bullet)}\right)
=e^{-i\theta\mathcal N_1}E_1
=
\frac{(3e^{i\theta}-1)e^{-i\theta\mathcal N_1}
-2e^{-i\theta\mathcal N_0}}
{3(e^{i\theta}-1)}.
\label{eq:initial-endpoint-identity}
}
Both $H_{\xx}$ and $\rho_c$ are translation invariant.
Translating every site by $-1$ therefore gives, for $x\geq1$,
\eq{
&\Tr_c\left[
E_{x+1}^{\epsilon}e^{i\theta\mathcal N_x}e^{-iH_{\xx}t}
e^{-i\theta\mathcal N_1}\rho_c e^{iH_{\xx}t}
\right]
=F_t^{(\varnothing,\epsilon)}(\theta;x-1).
\label{eq:shifted-initial-cut}
}
Combining Eqs.~\eqref{eq:initial-endpoint-identity} and \eqref{eq:shifted-initial-cut}, we obtain
\eqs{
F_t^{(\circ,\epsilon)}(\theta;x) +\frac{e^{-i\theta}}3F_t^{(\bullet,\epsilon)}(\theta;x)
&= 
\frac{(3e^{i\theta} -1) F_t^{(\varnothing,\epsilon)}(\theta;x-1) - 2F_t^{(\varnothing,\epsilon)}(\theta;x) }{3(e^{i\theta} -1)}
\\
&= (\mathsf T_{\theta} F^{(\varnothing,\epsilon)}_t(\theta; \cdot)) (x-1)
}
By Eq.~\eqref{eq:final-endpoint-reduction}, we find
\eq{
F_t^{(\circ,\epsilon)}(\theta;x) +\frac{e^{-i\theta}}3F_t^{(\bullet,\epsilon)}(\theta;x)
=
\bigl(\mathsf T_\theta^{1+\epsilon}F_t(\theta;\cdot)\bigr)(x-1).
\label{eq:merged-endpoint-reduction}
}
Taking Fourier coefficients proves Eqs.~\eqref{eq:bulk-empty-p} and \eqref{eq:bulk-merged-p}.
The factors $e^{-i\theta}$ shift the Fourier index from $n$ to $n+1$.

Only $F_t$ remains to be evaluated.
By the Jordan--Wigner transformation, $H_{\xx}$ is mapped to a noninteracting
fermion Hamiltonian, and an application of the Klich formula~\cite{Klich_2003,Schonhammer_2007}
expresses Eq.~\eqref{eq:basic-charge-trace} as a determinant.

To evaluate this determinant, we temporarily restrict the XX chain to the finite lattice
$\{-\Lambda+1,\ldots,\Lambda\}$ with open boundary conditions.
We denote the corresponding finite-size generating function by
$F_{t,\Lambda}(\theta;x)$.
In the following finite-size calculation, $H_{\xx}$ and $\rho_c$ are understood as their restrictions to this lattice.

Using charge conservation,
\eq{
F_{t,\Lambda}(\theta;x)
=
\Tr_c\left[
e^{-i\theta\overline{\mathcal N}_x}e^{-iH_{\xx}t}
e^{i\theta\overline{\mathcal N}_0}\rho_c e^{iH_{\xx}t}
\right],
}
where
\eq{
\overline{\mathcal N}_x:=\sum_{j=x+1}^{\Lambda}n_j.
}
Introducing
\eq{
\tilde{\rho}_\theta
:=
\bigotimes_{j=-\Lambda+1}^0
\left[\frac14\ket{\circ}\bra{\circ}+\frac34\ket{\bullet}\bra{\bullet}\right]_j
\bigotimes_{j=1}^{\Lambda}
\left[
\frac{1}{1+3e^{i\theta}}\ket{\circ}\bra{\circ}
+\frac{3e^{i\theta}}{1+3e^{i\theta}}\ket{\bullet}\bra{\bullet}
\right]_j,
}
we have
\eq{
F_{t,\Lambda}(\theta;x)
=
g(\theta)^{\Lambda}
\Tr_c\left[
e^{-i\theta\overline{\mathcal N}_x}e^{-iH_{\xx}t}
\tilde{\rho}_\theta e^{iH_{\xx}t}
\right],
}
where $g(\theta)$ is defined in Eq.~\eqref{eq:main-g-omega}.
Since $\tilde{\rho}_\theta$ is Gaussian, the Klich formula~\cite{Klich_2003, Schonhammer_2007} gives
\eq{
F_{t,\Lambda}(\theta;x) = g(\theta)^{\Lambda} \det\left[ \delta_{j,k}+(e^{-i\theta}-1)C_{j,k}(t) \right]_{j,k=x+1}^{\Lambda}.
\label{eq:F-det-klich}
}

Here, $C_{j,k}(t)$ is the single-particle correlation matrix associated with
$\tilde{\rho}_\theta$.
It is completely characterized by
\eq{
i\frac{d}{dt}C_{j,k}(t) = \frac12\left[ C_{j+1,k}(t)+C_{j-1,k}(t) -C_{j,k+1}(t)-C_{j,k-1}(t) \right],
\label{eq:evo-two-corr}
}
with open boundary conditions
\eq{
C_{-\Lambda,k}(t)=C_{j,-\Lambda}(t)=C_{\Lambda+1,k}(t)=C_{j,\Lambda+1}(t)=0,
}
and initial condition
\eq{
C_{j,k}(0)
=
\delta_{j,k}
\left[
\frac{3e^{i\theta}}{1+3e^{i\theta}}\ind{j>0}
+\frac34\ind{j\leq0}
\right].
}
Since Eq.~\eqref{eq:evo-two-corr} is linear and $\delta_{j,k}$ is a stationary solution,
\eq{
C_{j,k}(t) = \frac{3e^{i\theta}}{1+3e^{i\theta}}\delta_{j,k} + \left( \frac34-\frac{3e^{i\theta}}{1+3e^{i\theta}} \right)D_{j,k}(t),
\label{eq:C_jk-to-D_jk}
}
where $D_{j,k}(t)$ satisfies the same equation of motion with a domain wall initial condition,
\eq{
D_{j,k}(0)=\delta_{j,k}\ind{j\leq0}.
}
Substituting Eq.~\eqref{eq:C_jk-to-D_jk} into Eq.~\eqref{eq:F-det-klich}, we obtain
\eq{
F_{t,\Lambda}(\theta;x) = g(\theta)^x \det\left[ \delta_{j,k}+\omega(\theta)D_{j,k}(t) \right]_{j,k=x+1}^{\Lambda},
\label{eq:app-F-basic}
}
where $\omega(\theta)$ is defined in Eq.~\eqref{eq:main-g-omega}.

To solve the above problem, define $U_{j,k}(t)$ by
\eq{
i\frac{d}{dt}U_{j,k}(t)
=
\frac12\left[U_{j+1,k}(t)+U_{j-1,k}(t)\right],
\qquad
U_{j,k}(0)=\delta_{j,k},
\label{eq:app-U-definition}
}
with open boundary conditions
\eq{
U_{-\Lambda,k}(t)=U_{\Lambda+1,k}(t)=0.
}
The matrix $D_{j,k}(t)$ can then be written as
\eq{
D_{j,k}(t) = \sum_{\ell=-\Lambda+1}^{0}U_{j,\ell}(t)U_{k,\ell}(t)^*.
\label{eq:app-D-U}
}

Taking $\Lambda\to\infty$, Eq.~\eqref{eq:app-U-definition} gives
\eq{
U_{j,k}(t)=(-i)^{j-k}J_{j-k}(t).
\label{eq:app-U-bessel}
}
Consequently,
\eq{
D_{j,k}(t) = (-i)^{j-k} \sum_{\ell=0}^{\infty} J_{j+\ell}(t)J_{k+\ell}(t).
}
Using the Bessel recurrence relations, the infinite sum is evaluated exactly as
\eq{
K_{\mathrm{dBes}}(j,k;t)
:=\sum_{\ell=0}^{\infty}J_{j+\ell}(t)J_{k+\ell}(t)= \frac{t}{2(j-k)}[J_{j-1}(t)J_k(t) - J_{j}(t)J_{k-1}(t)].
\label{eq:app-discrete-bessel}
}
Thus,
\eq{
D_{j,k}(t)=(-i)^{j-k}K_{\mathrm{dBes}}(j,k;t),
}
where $K_{\mathrm{dBes}}$ is the discrete Bessel kernel defined in Eq.~\eqref{eq:app-discrete-bessel}.
Removing the phase $(-i)^{j-k}$ in Eq.~\eqref{eq:app-F-basic}
by the diagonal similarity transformation with entries $(-i)^j$ gives, in the infinite-system limit,
\eq{
F_t(\theta;x)=g(\theta)^x \det\left[\delta_{j,k}+\omega(\theta)K_{\mathrm{dBes}}(j,k;t)\right]_{j,k>x}.
\label{eq:app-discrete-bessel-det}
}
Using the determinant identity derived in Ref.~\cite{Moriya_2019},
\eq{
\det\left[\delta_{j,k}+\omega K_{\mathrm{dBes}}(j,k;t)\right]_{j,k>x}
=
\det\left[1+\omega K_{\mathrm{Bes}}^{(x)}\right]_{L^2([0,t^2])},
\label{eq:app-bessel-determinant-identity}
}
we obtain the continuous Bessel representation in Eq.~\eqref{eq:main-charge-determinant}.
Together with the endpoint reduction above, this evaluates all charge
factors needed in Eq.~\eqref{eq:main-exact-mgf} from one basic determinant.

\section[Long-time charge-factor asymptotics]
{Long-time asymptotics of $\mathcal P_t^{(a)}(x,n)$}
\label{app:charge-asymptotics}

Here, we derive the asymptotic form of $\mathcal P_t^{(a)}(x,n)$ used in
Eq.~\eqref{eq:charge-general-asymptotic}. The result is
\eq{
\mathcal P_t^{(a)}(x,n)
\simeq
\frac{2^{1-a}}{\sqrt{3t}}
\exp\left[-\frac{4\pi}{3t}\left(n-\frac{3x}{4}\right)^2\right],
\qquad a=0,1,2,\qquad x,n=\order{\sqrt t}.
\label{eq:app-charge-result}
}

We first consider the $a=0$ case.
From Eqs.~\eqref{eq:main-charge-determinant} and \eqref{eq:main-pt},
\eq{
\mathcal P_t^{(0)}(x,n)
=\int_{-\pi}^{\pi}\frac{\dd\theta}{2\pi}
e^{-in\theta}g(\theta)^x
\det\Big[1+\omega(\theta)K_{\mathrm{Bes}}^{(x)}\Big]_{L^2([0,t^2])}.
\label{eq:app-pt-0}
}

We first identify the region of $\theta$ that contributes to the leading asymptotics.
The Bessel kernel on $L^2([0,\infty))$ is an orthogonal projection, so its restriction to $[0,t^2]$ satisfies
\eq{
0 \leq \Tr \Big[\big(K^{(x)}_{\mathrm{Bes}}\big)^m \Big]_{[0,t^2]} 
\leq 
\Tr \Big[K^{(x)}_{\mathrm{Bes}}\Big]_{[0,t^2]}, \qquad m \geq1.
}
For real $\theta$, we have
\eq{
|g(\theta)|\leq1,
\qquad
\omega(\theta)=-\frac34\sin^2\frac{\theta}{2}.
}
From these relations, it follows that
\eqs{
\left| g(\theta)^x \det\Big[1+\omega(\theta)K_{\mathrm{Bes}}^{(x)}\Big]_{[0,t^2]} \right|
&\leq \left| \exp\left[\log \det \Big[1+\omega(\theta)K_{\mathrm{Bes}}^{(x)}\Big]_{[0,t^2]}\right] \right|
\notag
\\
&= \left| \exp\left[ \Tr \Big[\log (1+\omega(\theta)K_{\mathrm{Bes}}^{(x)} )\Big]_{[0,t^2]} \right]\right|
\notag
\\
& \leq \exp\left[ -\frac34\sin^2\frac{\theta}{2}\, \Tr \Big[K_{\mathrm{Bes}}^{(x)} \Big]_{[0,t^2]}\right].
\label{eq:app-charge-integrand-bound}
}

To evaluate the trace, we use the diagonal of the Bessel kernel:
\eq{
\Tr K_{\mathrm{Bes}}^{(x)}
&=\frac14\int_0^{t^2}\dd u\,
\left[
J_x(\sqrt u)^2-J_{x-1}(\sqrt u)J_{x+1}(\sqrt u)
\right]
\\
&=\frac12\int_0^t\dd r\,r
\left[
J_x(r)^2-J_{x-1}(r)J_{x+1}(r)
\right].
}
Rescaling $r=ts$, we obtain
\eq{
\Tr K_{\mathrm{Bes}}^{(x)}
=\frac{t^2}{2}\int_0^1\dd s\,s
\left[
J_x(ts)^2-J_{x-1}(ts)J_{x+1}(ts)
\right].
}

Away from the turning point $ts=x$, the Bessel functions are
exponentially suppressed for $ts<x$ and have oscillatory asymptotics for $ts>x$.
The turning-point region contributes only $o(t)$ to the trace.
Consequently, the leading contribution comes from the oscillatory region.
For fixed $s>0$ and $x=\order{\sqrt t}$, the oscillatory asymptotic
forms give
\eqnn{
J_x(ts)=\sqrt{\frac{2}{\pi ts}}\cos\Phi_x(ts)+o(t^{-1/2}),
\qquad
J_{x\pm1}(ts)=\pm\sqrt{\frac{2}{\pi ts}}\sin\Phi_x(ts)+o(t^{-1/2}),
}
where
\eqnn{
\Phi_x(r)=\sqrt{r^2-x^2}-x\arccos\frac{x}{r}-\frac{\pi}{4}.
}
The oscillatory terms cancel, yielding
\eqnn{
J_x(ts)^2-J_{x-1}(ts)J_{x+1}(ts)
=\frac{2}{\pi ts}+o(t^{-1}).
}
Integrating this leading term, with the endpoint and turning-point regions contributing only $o(t)$, we obtain
\eq{
\Tr K_{\mathrm{Bes}}^{(x)}
=\frac{t^2}{2}\int_0^1\dd s\,s\,\frac{2}{\pi ts}+o(t)
=\frac{t}{\pi}+o(t).
\label{eq:app-bessel-trace-asymptotic}
}

Thus, combining Eqs.~\eqref{eq:app-charge-integrand-bound}
and \eqref{eq:app-bessel-trace-asymptotic}, we find that the
leading contribution to the integral comes from
$\theta=\order{t^{-1/2}}$.
On this scale, we expand
\eq{
\log g(\theta)
=\frac{3i}{4}\theta-\frac{3}{32}\theta^2+\order{\theta^3},
\qquad
\omega(\theta)
=-\frac{3}{16}\theta^2+\order{\theta^4}.
}
Since
$0\leq\Tr[(K_{\mathrm{Bes}}^{(x)})^m]
\leq\Tr K_{\mathrm{Bes}}^{(x)}=\order{t}$,
the higher-order terms in the logarithm of the determinant satisfy
\eq{
\log\det\Big[1+\omega(\theta)K_{\mathrm{Bes}}^{(x)}\Big]
=
\omega(\theta)\Tr K_{\mathrm{Bes}}^{(x)}
+\order{t|\omega(\theta)|^2}
=
-\frac{3t}{16\pi}\theta^2+o(1).
}
Together with $x=\order{\sqrt t}$, this gives
\eq{
-in\theta+x\log g(\theta)
+\log\det\Big[1+\omega(\theta)K_{\mathrm{Bes}}^{(x)}\Big]
=
-i\left(n-\frac{3x}{4}\right)\theta
-\frac{3t}{16\pi}\theta^2+o(1).
}
Evaluating the resulting Gaussian integral, we obtain
\eq{
\mathcal P_t^{(0)}(x,n)
&\simeq
\int_{-\infty}^{\infty}\frac{\dd\theta}{2\pi}
\exp\left[
-i\left(n-\frac{3x}{4}\right)\theta
-\frac{3t}{16\pi}\theta^2
\right]
\\
&=
\frac{2}{\sqrt{3t}}
\exp\left[
-\frac{4\pi}{3t}\left(n-\frac{3x}{4}\right)^2
\right].
\label{eq:app-charge-a0}
}

We next consider $\mathcal P_t^{(1)}(x,n)$, which is given by
\eq{
\mathcal{P}_t^{(1)}(x,n) = \int_{-\pi }^\pi  \frac{\dd\theta}{2\pi }  e^{-in \theta }
\frac{(3e^{i\theta} -1) F_t(\theta;x) - 2 F_t(\theta;x+1)}{3 (e^{i \theta} -1)}
}
Writing
\eq{
d_t(\theta;x):=
\det\Big[1+\omega(\theta)K_{\mathrm{Bes}}^{(x)}\Big],
\qquad
F_t(\theta;x)=g(\theta)^x d_t(\theta;x),
}
one can rewrite the above equation as
\eq{
\mathcal{P}_t^{(1)}(x,n) =
\int_{-\pi }^\pi  \frac{\dd\theta}{2\pi }  e^{-in \theta } g(\theta)^x
\left[
\frac12d_t(\theta;x) -\frac{2g(\theta)}{3(e^{i\theta}-1)} \bigl\{d_t(\theta;x+1)-d_t(\theta;x)\bigr\}
\right].
\label{eq:app-pt-1}
}
Note that the expression is similar to that of $\mathcal{P}_t^{(0)}$ in Eq.~\eqref{eq:app-pt-0}.
Hence, by a similar argument, one sees that a leading contribution comes from the region $\theta = \order{t^{-1/2}}$.

To estimate the difference $d_t(\theta;x+1)-d_t(\theta;x)$,
we use the equivalent discrete Bessel determinant in Eq.~\eqref{eq:app-discrete-bessel-det}.
Since increasing $x$ by one removes the first row and column,
the cofactor formula for the inverse matrix gives
\eq{
\frac{d_t(\theta;x+1)}{d_t(\theta;x)}
&=
\left[
\bigl(I+\omega(\theta)K_{\mathrm{dBes}}\bigr)^{-1}
\right]_{x+1,x+1}
\\
&=
1-\omega(\theta)K_{\mathrm{dBes}}(x+1,x+1;t) +\order{\omega(\theta)^2}.
}
Hence,
\eq{
\frac{d_t(\theta;x+1)-d_t(\theta;x)}{d_t(\theta;x)} = -\omega(\theta)K_{\mathrm{dBes}}(x+1,x+1;t) +\order{\omega(\theta)^2}
=\order{\theta^2},
}
where we used $0\leq K_{\mathrm{dBes}}(x+1,x+1;t)\leq1$ and $\omega(\theta)=\order{\theta^2}$.
Thus, on the scale $\theta=\order{t^{-1/2}}$, the second term in Eq.~\eqref{eq:app-pt-1} is subleading.
The same Gaussian integration as in the $a=0$ case therefore gives
\eq{
\mathcal P_t^{(1)}(x,n) \simeq \frac{1}{\sqrt{3t}}
\exp\left[
-\frac{4\pi}{3t}\left(n-\frac{3x}{4}\right)^2
\right].
}

Finally, $\mathcal P_t^{(2)}(x,n)$ is evaluated in the same way.
Writing
$\Delta d_t(\theta;x)=d_t(\theta;x+1)-d_t(\theta;x)$,
we have
\eqnn{
\mathcal P^{(2)}_t(x,n)
= \int_{-\pi}^{\pi}\frac{\dd \theta}{2\pi} e^{-in \theta } g(\theta)^x\left[
\frac14d_t(\theta;x)
-\frac{2g(\theta)}{3(e^{i\theta}-1)}\Delta d_t(\theta;x)
+\frac{4g(\theta)^2}{9(e^{i\theta}-1)^2}\Delta^2d_t(\theta;x)
\right].
}
The first difference has already been estimated above.
Applying the same determinant-ratio expansion at $x$ and $x+1$ gives
\eq{
\frac{\Delta^2d_t(\theta;x)}{d_t(\theta;x)}
&=
\omega(\theta)\big[
K_{\mathrm{dBes}}(x+1,x+1;t)
-K_{\mathrm{dBes}}(x+2,x+2;t)
\big]
+\order{\omega(\theta)^2}
\\
&=
\omega(\theta)J_{x+1}(t)^2+\order{\omega(\theta)^2}
=\order{\theta^2/t+\theta^4},
}
where $J_{x+1}(t)^2=\order{t^{-1}}$ for $x=\order{\sqrt t}$.
Thus, both difference terms are subleading on the scale
$\theta=\order{t^{-1/2}}$, even after division by the corresponding
powers of $e^{i\theta}-1$.
The Gaussian integration therefore yields
\eq{
\mathcal P_t^{(2)}(x,n)
\simeq
\frac{1}{2\sqrt{3t}}
\exp\left[
-\frac{4\pi}{3t}\left(n-\frac{3x}{4}\right)^2
\right].
}

\bibliography{ref}
\end{document}